\documentclass[aps,  prd,twocolumn,  superscriptaddress]{revtex4-2}
\usepackage[english]{babel}

\usepackage{amsmath}
\usepackage{graphicx}
\usepackage[colorlinks=true, allcolors=blue]{hyperref}
\usepackage{dcolumn}
\usepackage{lineno}
\usepackage{multirow}
\usepackage{subfigure}
\usepackage{soul}

\begin{document}

\title{
First attempt of directionality reconstruction for atmospheric neutrinos in a large homogeneous liquid scintillator detector}

\newcommand{\SDU}{Shandong University, Jinan, China, and Key Laboratory of Particle Physics and Particle Irradiation of Ministry of Education, Shandong University, Qingdao 266237, China}
\newcommand{\IHEP}{Institute of High Energy Physics, Beijing 100049, China}
\newcommand{\UCAS}{School of Physical Sciences, University of Chinese Academy of Science, Beijing 100049, China}

\author{Zekun Yang}
\altaffiliation{Zekun Yang and Jiaxi Liu contributed equally to this work.}
\affiliation{\SDU}
\author{Jiaxi Liu}
\altaffiliation{Zekun Yang and Jiaxi Liu contributed equally to this work.}
\affiliation{\IHEP}
\affiliation{\UCAS}
\author{Hongyue Duyang}
\email{Corresponding author: duyang@sdu.edu.cn}
\affiliation{\SDU}
\author{Wanlei Guo}
\affiliation{\IHEP}
\author{Xinhai He}
\affiliation{\IHEP}
\affiliation{\UCAS}
\author{Yuxiang Hu}
\affiliation{\IHEP}
\affiliation{\UCAS}
\author{Teng Li}
\email{Corresponding author: tengli@sdu.edu.cn}
\affiliation{\SDU}
\author{Zhen Liu}
\email{Corresponding author: liuzhen@ihep.ac.cn}
\affiliation{\IHEP}
\author{Wuming Luo}
\email{Corresponding author: luowm@ihep.ac.cn}
\affiliation{\IHEP}
\author{Xiaojie Luo}
\affiliation{\IHEP}
\affiliation{\UCAS}
\author{Wing Yan Ma}
\affiliation{\SDU}
\author{Xiaohan Tan}
\affiliation{\SDU}
\author{Fanrui Zeng}
\affiliation{\SDU}
\author{Yongpeng Zhang}
\affiliation{\IHEP}

%\linenumbers
\begin{abstract}
The directionality information of incoming neutrinos is essential to atmospheric neutrino oscillation analysis since it is directly related to the oscillation baseline length. 
Large homogeneous liquid scintillator detectors, while offering excellent energy resolution, are traditionally very limited in their capabilities of measuring event directionality. 
In this paper, we present a novel directionality reconstruction method for atmospheric neutrino events in large homogeneous liquid scintillator detectors based on waveform analysis and machine learning techniques.
We demonstrate for the first time that such detectors can achieve good direction resolution and potentially play an important role in future atmospheric neutrino oscillation measurements. 
\end{abstract}
\maketitle

\section{Introduction}

Liquid scintillator (LS) detectors play an important role in neutrino physics.
They typically offer low-threshold and high-precision energy measurements, ideal for low-energy topics such as reactor neutrinos and solar neutrinos. 
Notable examples include the first reactor neutrino oscillation measurement by KamLAND \cite{KamLand}, sub-MeV solar neutrino detection by Borexino \cite{Borexino, BOREXINO:2018ohr}, and the $\theta_{13}$ measurement by Daya Bay \cite{DayaBay:2012fng}. 
The central detector (CD) of Jiangmen Underground Neutrino Observatory (JUNO) that is currently under construction in China will be the largest detector of this kind \cite{JUNO, JUNO_phydet}. 
With 20\,kton target mass and 78\% photo-multiplier tube (PMT) coverage, JUNO is designed to determine the neutrino mass ordering (NMO) via a precise measurement of the oscillation spectrum of reactor neutrinos. 

Atmospheric neutrino measurements led to the discovery of neutrino oscillations in the late 1990s \cite{Super-Kamiokande:1998kpq},
and are expected to continue contributing to the knowledge of neutrino mixing angles, NMO, and the CP violation phase in the next decades. 
Event directionality information is mandatory to atmospheric neutrino oscillation measurements since the oscillation baseline length varies as a function of the neutrino zenith angle.
Directionality measurement in large homogeneous LS detectors, however, is very challenging. 
On one hand, unlike tracking detectors such as liquid argon time projection chambers used by DUNE \cite{DUNE:2016hlj}, LS detectors do not offer direct track information. 
On the other hand,
Cherenkov light, while offering excellent directional information in water detectors such as Super-K \cite{Super-Kamiokande:1998kpq}, 
is only about a few percent of scintillation light in a typical LS detector.
There have been efforts on separating Cherenkov light from scintillation light by utilizing slow scintillators \cite{Wei:2016vjd,Wang:2017etb} or photo sensors with high timing resolution \cite{Caravaca:2016fjg,Caravaca:2016ryf}. 
A notable recent progress of these kinds is the reconstruction of solar neutrino's directionality by SNO+ \cite{SNO+:2023}.
Water-based scintillator technique with tunable Cherenkov light to scintillation light ratio also shows great potential \cite{Yeh:2011zz}.
While most of these techniques are developed for lower energy topics such as solar neutrinos, they certainly can be adapted for atmospheric neutrino oscillation measurements. 
However, major hardware upgrades are required for those techniques to be applicable to existing detectors.   

In recent years, new ideas have been developed which turn to scintillation light for directionality information \cite{Learned:2009rv}. 
In the low energy (tens of MeV or below) region, a charged particle deposits energy in LS like a point source, leading to an isotropic scintillation light distribution. 
But at higher energies, the space and time distributions of scintillation photons in the detector are the superposition of light from points along the particle tracks, the characteristics of which therefore reflect the event topology in the detector. 
A topological reconstruction method is developed based upon this concept and shows good potential in reconstructing muon events with the assumption of a known reference point \cite{Wonsak:2018uby}. 
But the reconstruction performance of more complex events like those from atmospheric neutrinos has never been reported. 

In this paper, we present a novel method of event directionality reconstruction for atmospheric neutrinos in large homogeneous LS detectors based on waveform analysis and machine learning (ML) techniques with information from scintillation light. 
This method extracts features relevant to event directionality from PMT waveforms, and uses them as inputs to ML models which are trained to predict the incoming neutrino direction. 
Performances on atmospheric neutrino events using Monte Carlo (MC) simulation with different ML models are discussed.
It is demonstrated that a large homogeneous LS detector such as the JUNO CD can achieve good directionality resolution for atmospheric neutrino events, making it suitable for atmospheric neutrino oscillation measurements.
This is the first demonstration of atmospheric neutrino's directionality reconstruction in any homogeneous LS detectors with MC studies.

This paper is organized as follows. 
Section \ref{sec:meth} describes the idea and technique in detail. 
The virtual apparatus and data samples used in this study can be found in section \ref{sec:sim}. 
The algorithms for PMT waveform feature extraction 
and details of the ML models are discussed in section \ref{sec:fe} and \ref{sec:ml}.
The performances of models with MC simulation are presented in section  \ref{sec:rp}. 
In section \ref{sec:dis}, the advantage of this method by reconstructing the incident neutrino direction directly and the dependence on neutrino interaction models are discussed. Finally, section \ref{sec:so} summarizes the study.

\section{Methodology \label{sec:meth}}

\begin{figure}
\centering
\includegraphics[width=0.4\textwidth]{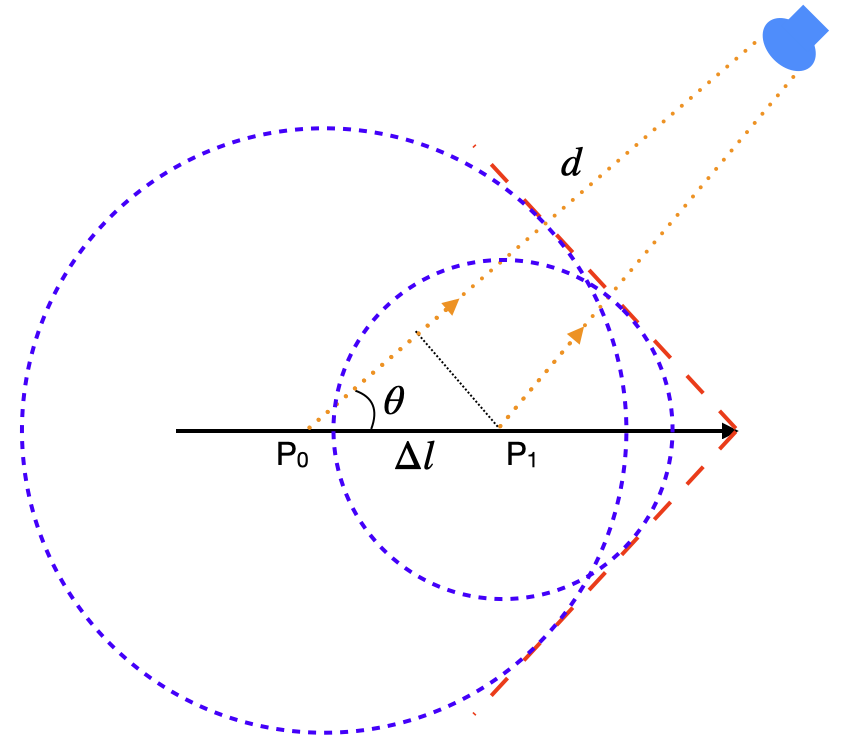}
\caption{\label{fig:track_pmt} Illustration of the scintillation light (orange dashed lines) from a charged particle track (black solid line) reaching a PMT. The isotropic emission of scintillation light from two points (P$_1$ and P$_2$, separated by a distance $\Delta l$) is illustrated by blue dashed circles. The scintillation-light front at a certain time forms a cone-like structure, illustrated by red dashed lines.}
\end{figure}

The light seen by PMTs of an LS detector is a superposition of light generated at many points on particle tracks inside the detector. 
If a particle travels with a speed faster than the speed of light in LS, scintillation light forms a cone-like front structure, as discussed in \cite{Learned:2009rv}.
For each event, the hit time of the earliest photon reaching a PMT (``first hit time'') therefore naturally offers information on the event directionality.

In addition to the first hit time, more information can be extracted from the hit time distribution of photons reaching PMTs. 
Consider two arbitrary points on a single charged particle track, P$_0$ and P$_1$, separated by a small distance $\Delta l$, as illustrated in Fig.~\ref{fig:track_pmt}. 
The time difference between the first scintillation photons from P$_0$ and P$_1$ arriving at a PMT at an angle $\theta$ with respect to the track direction is

\begin{equation}
    \Delta t = \left |\frac{ \Delta l }{v} - \frac{ \Delta l \cos{\theta}}{c/n}\right | = \Delta l \left |\frac{1- n\beta \cos{\theta}}{v}\right |
    \label{eqn:delta_t}
\end{equation}
where $v$ is the speed of the particle, $c$ is the speed of light in vacuum and $n$ is the effective group refraction index of LS. 
Here it is assumed that the distance from P$_{0}$ or P$_{1}$ to the PMT is much larger than $\Delta l$.
With very small $\Delta t$ and $\Delta l$, Eqn~\ref{eqn:delta_t} can be rewritten in a differential form:

\begin{equation}
    \frac{ \mathrm d l}{\mathrm d t} = \frac{v}{\left |1 - n\beta\cos{\theta}\right |} .
\end{equation}

\begin{figure}
\centering
\includegraphics[width=0.45\textwidth]{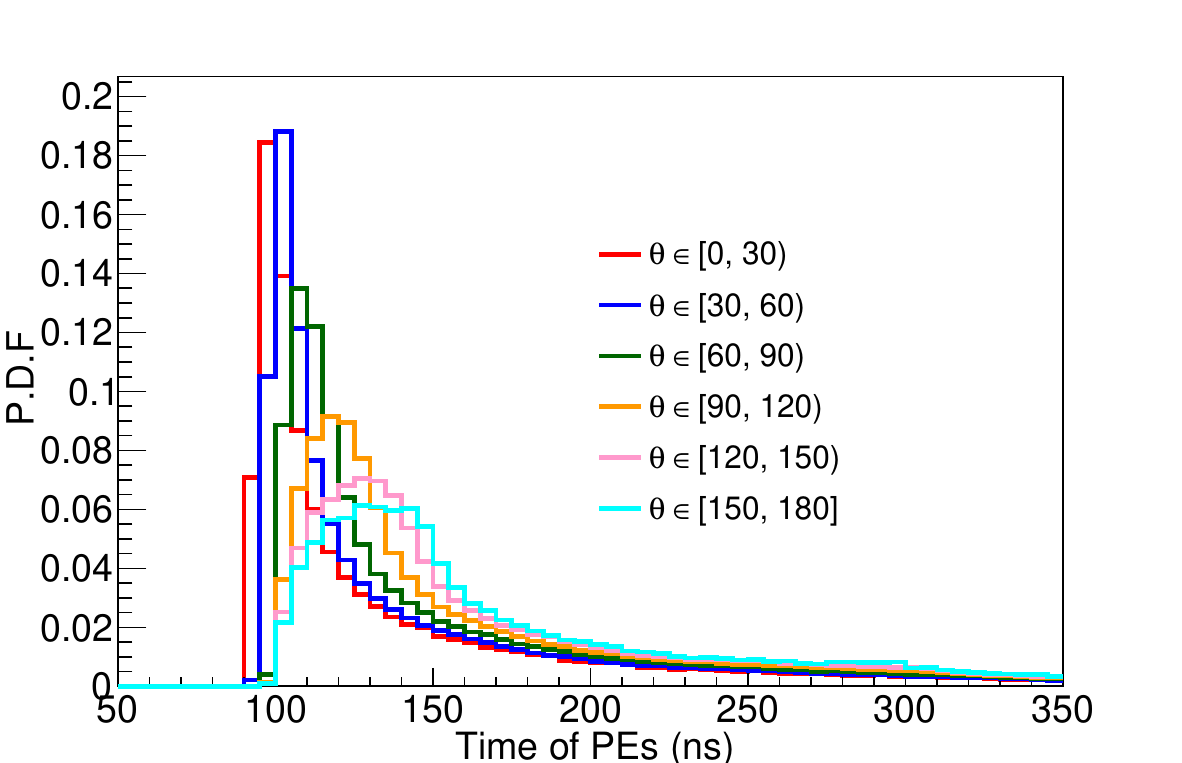}
\caption{\label{fig:npe_time} The normalized $N_{\text{PE},i}(t)$ distributions for PMTs at different angles with respect to the track by a 1\,GeV muon simulated at the center of the detector. The PMT angle $\theta$ is defined as the intermediate angle between the muon direction and the line connecting the middle point of the muon track and the PMT. Distinct $N_{\text{PE},i}(t)$ shapes can be observed for PMTs at different angles.}
\end{figure}

$\frac{\mathrm d l}{\mathrm d t}$ describes the change rate of track length $l$ that is visible to the PMT, which depends upon $\theta$.  
Since the amount of scintillation light emitted is related to $l$, 
it is easy to conclude that the number of photons received by a PMT as a function of time (light curve) also depends upon $\theta$. 
The exact shape of the light curve obviously also depends on the energy deposition along the track and LS properties.  
Interestingly $\frac{\mathrm  d l}{\mathrm d t}$ reaches its maximum when $\cos{\theta} = \frac{1}{n\beta}$ if $\beta > \frac{1}{n}$ (i.e., the particle speed is greater than the speed of light in LS), the same angle as Cherenkov radiation. 
A PMT at an angle closer to $\theta = \arccos (1/n\beta)$ sees a light curve with a much steeper front edge, while a PMT further away from this angle receives light more spreading out in time.  
This is then reflected in how the number of photoelectrons (PEs) of a PMT evolves as a function of time ($N_{\text{PE},i}(t)$ for the $i$-th PMT, Fig.~\ref{fig:npe_time} ), and finally in the PMT's waveform. 
In principle, PMT waveforms in an LS detector contain all the information one needs for a directionality reconstruction. 
In practice, however, it can be very difficult to extract directionality information directly from waveforms given the complex relationship between the two and considering the large number of PMTs involved.
A two-step method is developed to simplify the task. 

First, instead of using the full waveforms, key features are extracted from waveforms to keep only the useful information relevant to the event directionality. 
Examples of such features include the first hit time, the total charge, and the slope of the waveforms' front edge. 
Details of waveform features and their extraction are discussed in section \ref{sec:feature}.

While it is possible to figure out some simple relations between the waveform features and the directions for single particles with distinctive track-like topology such as cosmic muons, the task is still not as easy for atmospheric neutrinos in the GeV energy region. 
The photons detected by PMTs in atmospheric neutrino events may contain the contributions from both charged leptons and hadrons produced by neutrinos interacting with the detector nuclei.
Moreover, electrons or hadrons usually produce showers in LS, making it even more difficult to deduce the neutrino directionality from the PMT waveform features with any traditional methods such as likelihood-based methods.
To further simplify the task, a machine learning (ML) approach is developed.
ML  is now widely used in particle physics experiments and shows great potential in extracting information from detector signals and improving detector performances, making it a possible solution to this problem.  
In this study, several ML models are developed and trained with a large number of atmospheric neutrino events to find out the original neutrino direction from feature patterns. 
Note that in principle it is also possible to reconstruct the direction of the final-state charged lepton.
Neutrino direction is chosen as the ML model output because it has a larger impact on the neutrino oscillation sensitivity,  
while the relation between the charged lepton direction and the neutrino direction is smeared by differential neutrino-nucleus cross sections. Details of the ML models used in this study are discussed in section \ref{sec:ml}.

\section{Detector and Simulation \label{sec:sim}}

\begin{figure}
\centering
\includegraphics[width=0.4\textwidth]{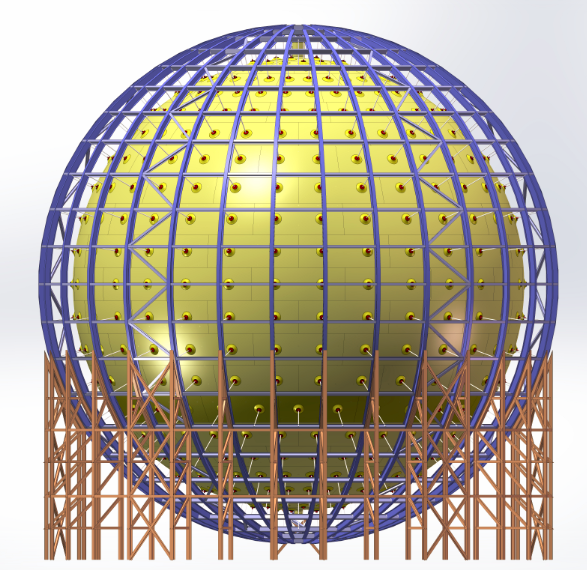}
\caption{\label{fig:junocd} Drawing of the JUNO central detector design. } 
\end{figure}

The simulation of JUNO CD is used to demonstrate the method. 
The JUNO CD is a large-scale spherical LS detector under construction (Fig.~\ref{fig:junocd}). 
Linear alkylbenzene is used as the detection medium, with 
2.5\,g/L 2,5-diphenyloxazole (PPO) as the fluor and 3 mg/L p-bis- (o-methylstyryl)-benzene (bis-MSB) as the wavelength shifter \cite{JUNO_phydet, JUNO_LS}. 
The total LS mass is 20\,kton, contained in an acrylic sphere with a radius of 17.7\,m.  
17,612 20-inch PMTs will be arranged facing-inward on a 19.5\,m radius spherical structure to detect photons, 
including 12,612 dynode PMTs and 5,000 MCP PMTs, 
with a total coverage of about 75\%.  
Parameters of the 20-inch PMTs used in the simulation are summarized in Tab.~\ref{table:pmt_par}. 
In addition, 25,600 3-inch PMTs will also be installed, although they are not used in this paper for simplicity.

\begin{table}
\centering
\caption{Summary of the PMT parameters used in the simulation.} 
\begin{tabular}{lcc}
 \hline
 \hline
PMT Type & Dynode & MCP \\
 \hline
Charge resolution (p.e.) & 0.28$\pm$0.02  & 0.33$\pm$0.03 \\
Time transit spread (ns) & 1.1$\pm$0.1  & 7.6$\pm$0.1\\ 
dark noise rate (kHz) & 15$\pm$6 & 32$\pm$16  \\
 \hline 
  \hline 
\end{tabular}
\label{table:pmt_par}
\end{table}

GENIE (v3.0.6) event generator \cite{Andreopoulos:2015wxa,Andreopoulos:2021prd} is used to simulate charged current (CC) neutrino interactions in the detector with neutrino flux from \cite{Honda:2015fha}. 
The detector response is simulated based on GEANT 4 \cite{GEANT4:2002zbu,Allison:2006ve},  
with a customized package developed for the simulation of various electronic effects. 
In the end, a simulated data sample with about 135k  $\nu_\mu$/$\bar{\nu}_{\mu}$‐CC and 57k $\nu_e$/$\bar{\nu}_{e}$‐CC  events with incoming neutrino energy between 1\,GeV and 20\,GeV are produced for the training and validation of ML models. 
In addition, an independent sample is simulated with the NuWro event generator \cite{Golan:2012rfa} to check the robustness of the performance with variations of neutrino interaction models.

\section{PMT waveform feature extraction  \label{sec:fe}}
\label{sec:feature}

\begin{figure}
\centering
\includegraphics[width=0.45\textwidth]{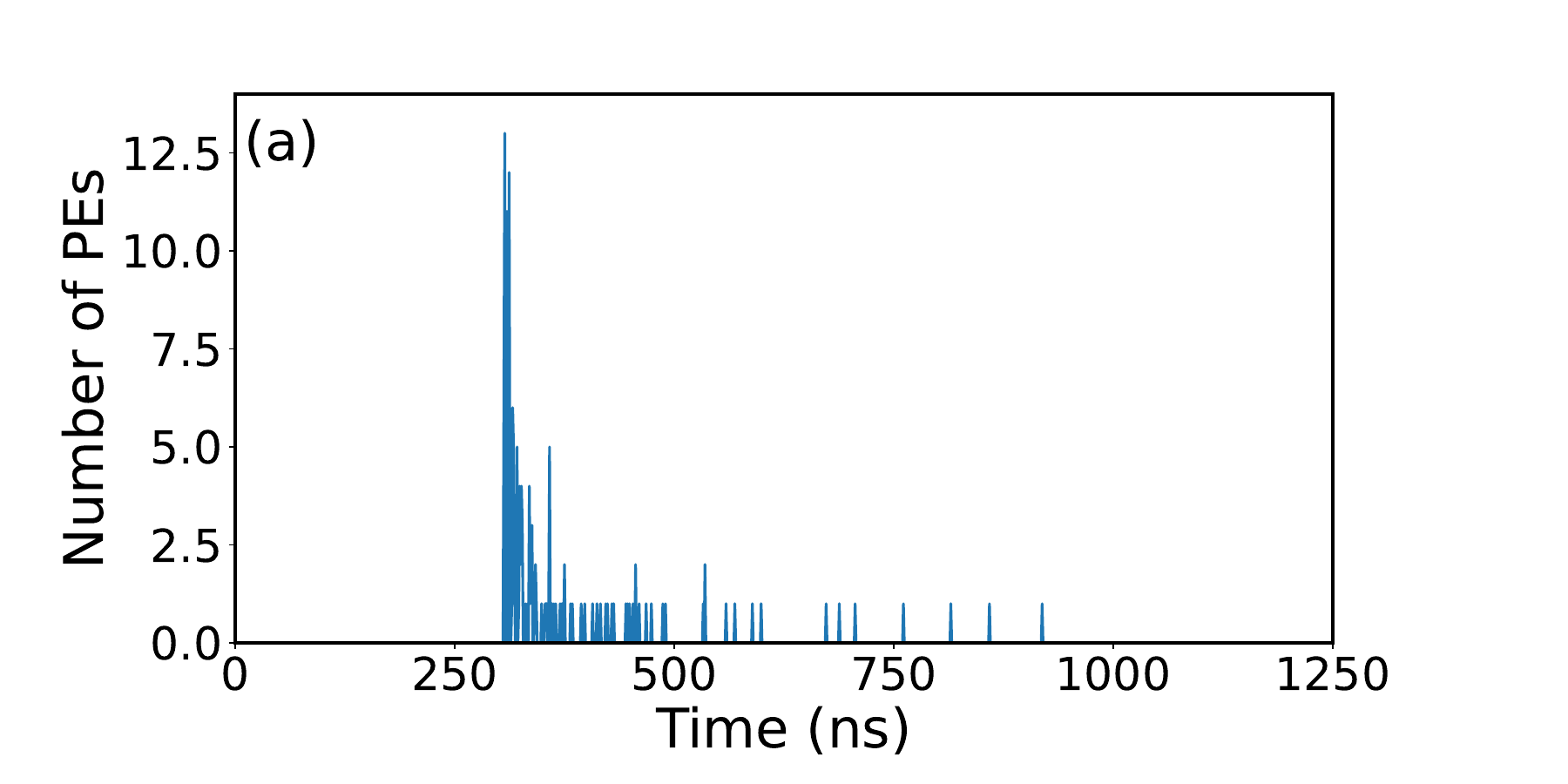}
\includegraphics[width=0.45\textwidth]{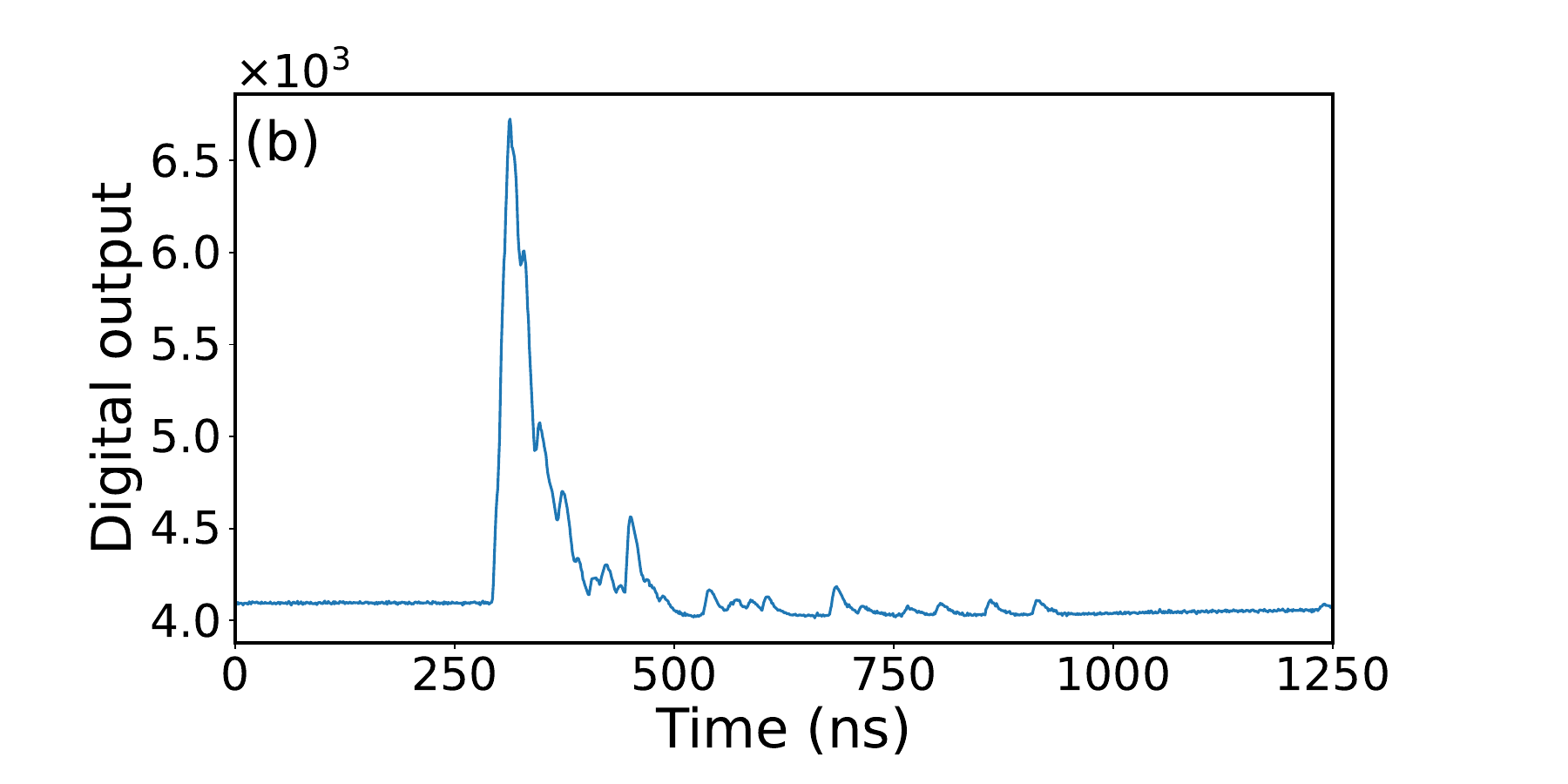}
\includegraphics[width=0.45\textwidth]{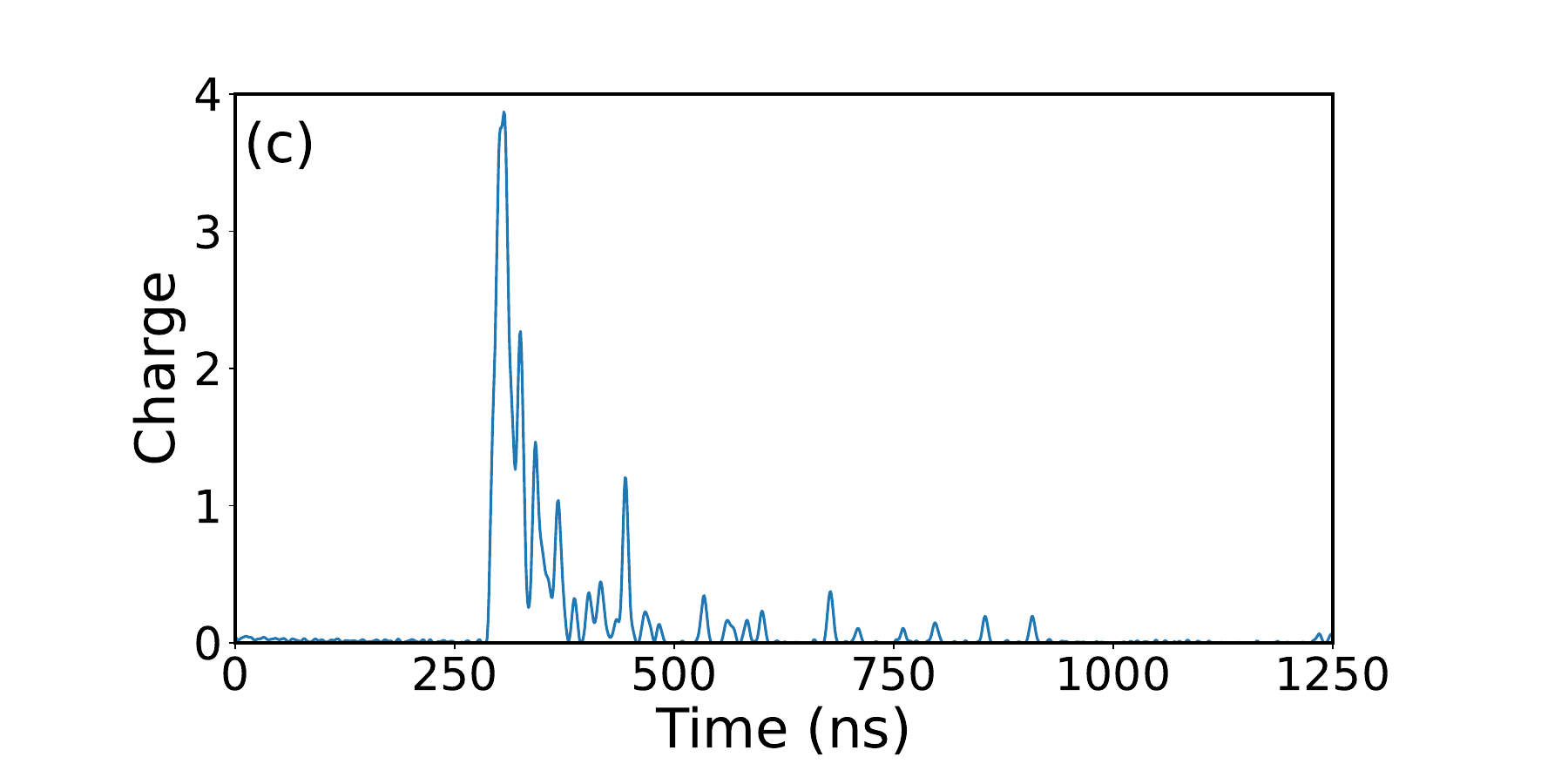}
\caption{The photoelectron signal (a), observed waveform (b), and the deconvoluted waveform with baseline adjusted (c) of one of the PMT.  The signal was produced by one of the atmospheric $\nu_{\mu}$ events in Monte Carlo simulation. }
\label{fig:waveform}
\end{figure}

The photoelectron signal collected by PMTs is converted to waveforms through Flash Analog-to-Digital converters modules. The observed PMT waveforms are affected by charge-smearing effects and various noise sources, which limits the accuracy of the event reconstruction. To improve the signal quality, the deconvolution method is used to mitigate the single photoelectron (SPE) waveform response of the PMTs and reduce the high-frequency noise \cite{deconv,HUANG201848}.

The deconvolution method works by estimating the SPE waveforms and noise properties of PMTs, and then using this information to reconstruct the photoelectron signal.
Consider the fact that the PMT waveform can be decomposed into two parts: signal and noise. The signal is a sum of individual PEs, with each one convoluted with the SPE waveform response. The time-dependent observed waveform of the $i_{th}$ PMT, $O_i(t)$, can be expressed as
\begin{equation}
O_i(t) = H_i(t) * U_i(t) + N_i(t)
    \label{eqn:observed_waveform}
\end{equation}
where $H_i(t)$ is the true photoelectron hit at time $t$, $U_i(t)$ is the SPE waveform, $N_i(t)$ represents the white noise, and $*$ denotes the convolution.

The deconvoluted waveform of the $i_{th}$ PMT, $H_i^{\text{deconv}}$, which estimates the true photoelectron distribution, can be calculated using the following formula in the frequency domain:

\begin{equation}
H_i^{\text{deconv}}(f)=[O_i(f)F_i(f)]/T_i(f)
    \label{eqn:reco_waveform}
\end{equation}
where $F_{i}(f)$ is the filter, and $T_{i}(f)$ is the SPE template in the frequency domain.
The SPE template for each PMT is obtained by averaging over a large number of SPE waveforms produced from a simulated $^{68}$Ge calibration sample.
The filter is calculated as $[(s(f)+n(f))^2 - n(f)^2]/(s(f)+n(f))^2$, where $s(f)+n(f)$ and $n(f)$ are the average amplitude of the SPE signal and the noise in the frequency domain, respectively.  
The final deconvoluted waveform $H_i^{\text{deconv}}(t)$ is obtained by converting $H_i^{\text{deconv}}(f)$ into the time domain with baseline adjusted by setting the average white noise level to zero.
Fig.~\ref{fig:waveform} shows the PE signal, observed waveform, and deconvoluted waveform for one of the PMTs as an example.

The key features of the waveform distributions are extracted from the deconvoluted waveform in the first 1.25 $\mu$s readout window of each PMT.
These features include:
 \begin{itemize} 
\item \textbf{Total charge}, which is calculated by integrating the charge over the entire readout time window;
 \item \textbf{First hit time (FHT)}, which is calculated by using a constant fraction discriminator method with a threshold of 20\% of peak charge; 
\item \textbf{Slope}, which describes the average slope of the deconvoluted waveform in the first 4\,ns after the first hit time:

 $ \text{slope}_{i} = [H_i^{\text{deconv}}(t = \text{FHT}+4) -H_i^{\text{deconv}}(t = \text{FHT})]/4 $

 \item \textbf{Charge ratio}, which is defined as the ratio of charge in the first 4\,ns after FHT to the total charge; 
\item \textbf{Peak charge} and \textbf{peak time}, which correspond to the charge and time of the peak of the deconvoluted waveform, respectively. 
 \end{itemize} 
In principle, additional features could be extracted to provide further details about the waveforms. Only the tested features that turn out to have a non-negligible impact on the directionality reconstruction are listed. 
Feature images of all the 20-inch PMTs are used as inputs to the ML models. 

\section{Machine learning models \label{sec:ml}}

The Convolutional Neural Network (CNN) \cite{cnn} technique is very powerful in processing images, and has been widely used in particle physics experiments. 
Features extracted from PMTs in an LS detector form image-like data on the PMT surface, which is well-suited for CNN models to deal with. 
However, given the PMTs in many LS detectors such as JUNO are arranged on a spherical surface, the features extracted cannot be fed to a CNN directly since typical models based upon CNN can only process images on the Euclidean domain and there is no way to define a sliding window (i.e. the convolution kernel) on the sphere.

Three different approaches are developed to deal with this problem.
The first approach projects spherical data on a planar surface so that it can easily be processed by various state-of-art ML models such as EfficientNetV2 \cite{EffiNetV2}. 
The second approach utilizes a model based upon DeepShere \cite{DeepSphere}, a graph convolution neural network (GCNN) designed to process spherical data dedicatedly. 
Lastly, a 3D model based upon PointNet++ ~\cite{PointNet2} which processes PMT data as a 3D point cloud is utilized. 
For each model, the output is designed to be the ($x$,$y$,$z$) components of the unit directional vector representing the incident neutrino direction. 
The loss function is defined as the Euclidean distance between the true ($x$,$y$,$z$) and the reconstructed ($x'$,$y'$,$z'$) points. 
This definition is completely rotation invariant and turns out to be able to avoid reconstruction bias while maintaining good resolution performance.
Details of the three kinds of models are discussed in the following subsections, and their performances are compared in section \ref{sec:rp}. 

\subsection{A planar machine learning model: EfficientNetV2 }
\label{sec:2d_model}

Many CNN models, such as VGG \cite{VGG}, ResNet \cite{ResNet}, EfficientNet \cite{EffiNet}, and EfficientNetV2 \cite{EffiNetV2}, have emerged with superior performance in the ImageNet Large-Scale Visual Recognition Challenge (ILSVRC). While Transformer-based architectures \cite{VIT,Swin} have shown promise in recent ILSVRC, they require significantly larger datasets for training. In contrast, CNNs' convolution operation extracts high-level and multi-level features from planar images, giving them superior generalization capabilities with small datasets.
 
%As mentioned, the features of each event form spherical image-like data. To adapt the data for planar models, it is projected onto a planar representation. The direction of the incident neutrino is subsequently reconstructed using the state-of-the-art CNN model, EfficientNetV2, known for its superior performance and reduced training time.

%A Tanh layer, appended after the fully connected (FC) layer of the original EfficientNetV2-S architecture \cite{EffiNetV2}, normalizes the output to fall within the -1 to 1 range. 
%To incorporate the planar machine learning models, the PMT map is projected onto a two-dimensional $\theta_{\text{PMT}}$‐$\phi_{\text{PMT}}$ grid (Fig.~\ref{fig:effnet_feature_2d}), where $\theta_{\text{PMT}}$ and $\phi_{\text{PMT}}$ are the zenith and azimuth angles of the PMT position. 
%The grid size of $128\times224$ is chosen to ensure each grid cell corresponds to at most one PMT. The PMT features are first filled into the $\theta_{\text{PMT}}$‐$\phi_{\text{PMT}}$ grids and then stacked together and fed into the models.

\begin{figure}
\centering
\includegraphics[width=0.45\textwidth]{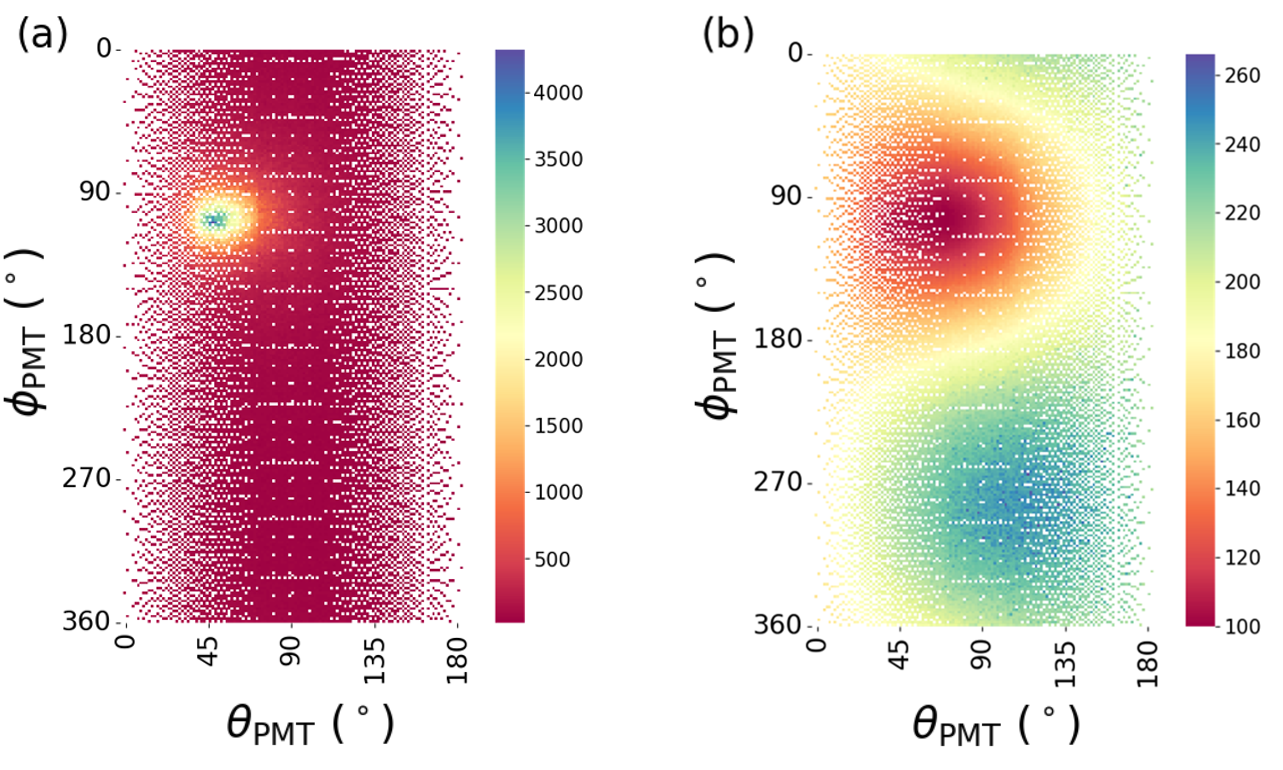}
\caption{The projected total charge (a) and first hit time (b) information on $\theta_{\text{PMT}}$‐$\phi_{\text{PMT}}$  grids for one of the atmospheric $\nu_\mu$‐CC events. }
\label{fig:effnet_feature_2d}
\end{figure}

As mentioned, the features of each event form spherical image-like data. 
To incorporate the planar machine learning models, the PMT map is projected onto a two-dimensional $\theta_{\text{PMT}}$‐$\phi_{\text{PMT}}$ grid (Fig.~\ref{fig:effnet_feature_2d}), where $\theta_{\text{PMT}}$ and $\phi_{\text{PMT}}$ are the zenith and azimuth angles of the PMT position. 
The grid size of $128\times224$ is chosen to ensure each grid cell corresponds to at most one PMT. 
The PMT features are first filled into the $\theta_{\text{PMT}}$‐$\phi_{\text{PMT}}$ grids and then stacked together. 
The direction of the incident neutrino is subsequently reconstructed using the state-of-the-art CNN model, EfficientNetV2, known for its superior performance and reduced training time.
A Tanh layer, appended after the fully connected (FC) layer of the original EfficientNetV2-S architecture \cite{EffiNetV2}, normalizes the output to fall within the -1 to 1 range. 

\subsection{A spherical machine learning model: DeepSphere}

DeepShere is a GCNN model originally developed for cosmology studies to deal with data distributed on a sphere \cite{DeepSphere}. 
One of the major advantages DeepShere provides is that it avoids projecting data to a planar surface, and maintains rotation co-variance, i.e. rotation of the input variables causes the same rotation of the predicted value. 
The main idea of DeepShere is to model the spherical data to a graph of connected pixels, and perform graph convolution based on spectral graph method. 

\begin{figure*}
\centering
\includegraphics[width=0.8\textwidth]{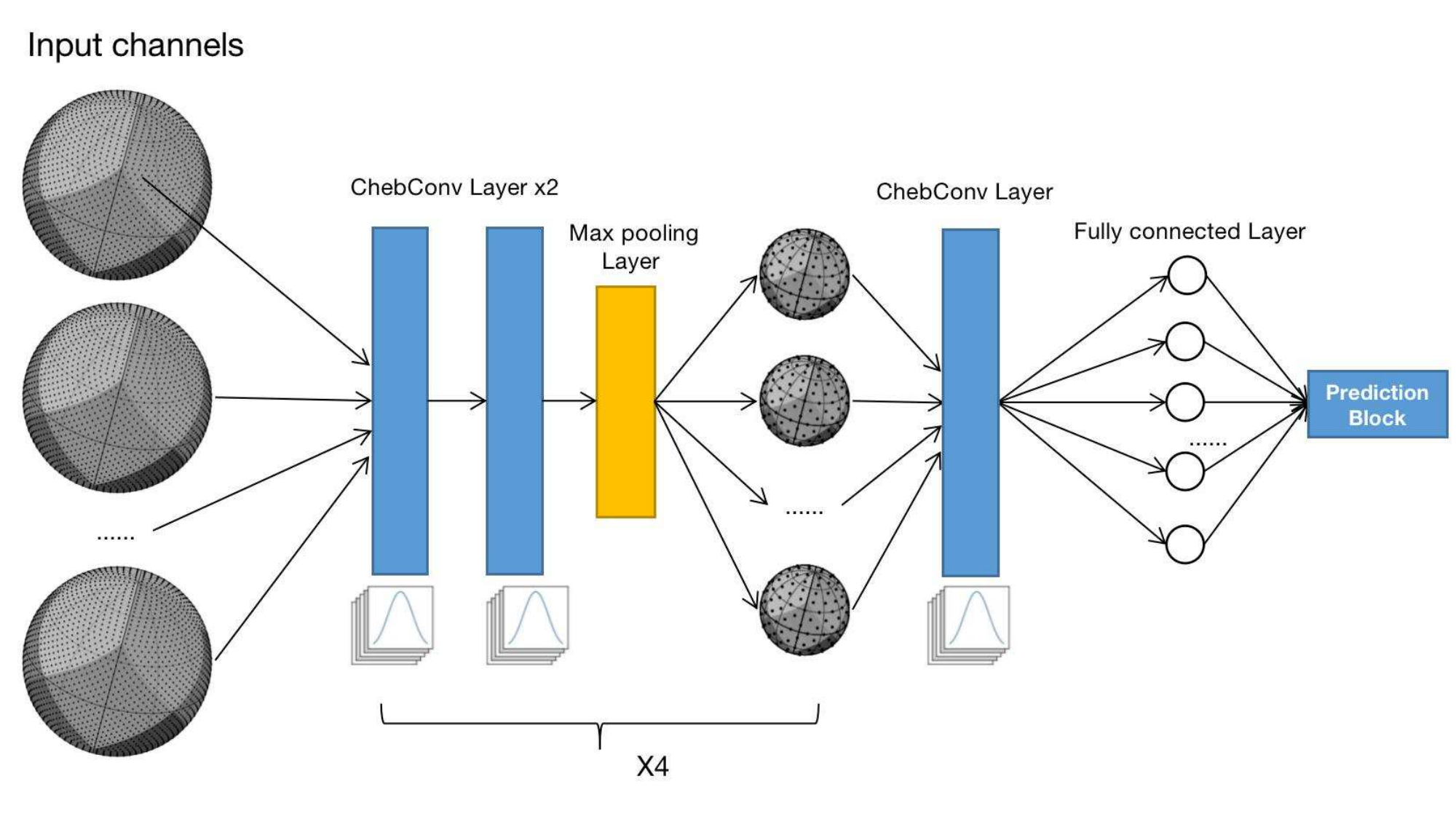}
\caption{\label{fig:ds_model} Architecture of the model based on DeepSphere. Subsequent to the input features are four sets of Chebyshev convolution layer and max pooling layer blocks (each with two Chebyshev convolution layers and one max pooling layer). Then another Chebyshev convolution layer and a fully connected layer are followed by the prediction block.}
\end{figure*}

To utilize DeepSphere, the spherical surface formed by PMTs is pixelized by the HEALPix scheme \cite{HEALPix}, which divides the surface by $N_{\text{pix}} = 12N_{\text{side}}^2$ equal-sized pixels. In this study, $N_{\text{side}} = 32 $ is used, so that the total pixel number is 12,288, which is less than the total PMT number. 
Since it is likely that one pixel covers more than one PMTs, the total charge of the pixel is the sum of all PMTs in the pixel, while FHT is taken as the earliest one. The other features are simply calculated by averaging PMTs in the pixel. 
Each pixel is then represented as a graph vertex before being fed into the model.
The connectivity between graph vertices is represented by the adjacency matrix $W_{i,j}$, which is calculated via 
 $\exp{(-\frac{\left\| \vec{x}_{i}-\vec{x}_{j} \right\|^{2}}{^{\rho^{2}}})}$
if vertex $\textit{i}$ and vertex $\textit{j}$ are neighbors (0 if otherwise), where $\vec{x}_{i}$ and $\vec{x}_{j}$ are the coordinates, $\rho$ is the averaged distance over all connected pixels.  
Since there is no straightforward way to implement convolution via sliding windows on the sphere, the convolution on the graph is implemented through an efficient spherical harmonic transform (SHT), proposed by \cite{cohen2018spherical,esteves2018learning}.
The architecture of the model developed based on DeepShere is shown in Fig.~\ref{fig:ds_model}. It consists of 4 sets of convolution blocks, followed by one Chebyshev convolution (ChebConv) layer, a fully connected layer and lastly a prediction block. 
Each convolution block contains two Chebyshev convolution layers and one max pooling layer.

\subsection{A 3D-based machine learning model: PointNet++}
Each PMT within the JUNO CD can be regarded as an individual discrete point. 
After the PMT waveform feature extraction, each event can be represented as a 3D point-cloud data, where each point contains the information of the three coordinates and the extracted features of one PMT.
Therefore it can be directly fed into 3D point-cloud-based machine learning models.

PointNet~\cite{PointNet} is one of the most influential 3D-based deep learning models, which directly takes point clouds as input. 
It is effective in learning global features from point cloud data but incapable of learning local neighborhood features. To overcome this issue, a hierarchical network based on PointNet, the so-called PointNet++~\cite{PointNet2}, was proposed. 
The PointNet++ architecture is designed to recursively sub-sample a small neighborhood from the whole point clouds, group the neighborhood into larger units, and then extract local features with mini-PointNet. Therefore, it is able to capture the fine-grained local features besides learning global ones.
 
In this study, the object classification version of PointNet++'s implementation~\cite{Pointnet2_code} is selected as the backbone network, and a tanh layer is added at the end of the network.
The input data are those featured by the Cartesian coordinates of the PMT position, together with the PMT features mentioned in section~\ref{sec:fe}.

\section{Performance \label{sec:rp}}

\begin{figure}
\centering
\includegraphics[width=0.4\textwidth]{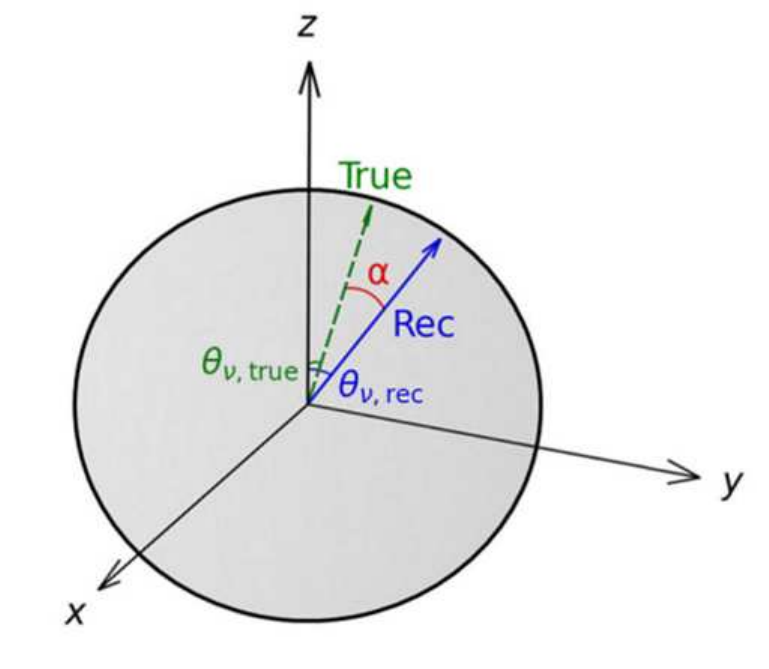}
\caption{\label{fig:angles} Illustration of the angles defined to benchmark the reconstruction performance: $\alpha$ is the angle between the true (green, dashed) and reconstructed (blue, solid) directional vector, $\theta_{\nu,\text{true}}$ and $\theta_{\nu,\text{rec}}$ are the zenith angle of the true and reconstructed vectors respectively.
}
\end{figure}

\begin{figure*}
\centering
\includegraphics[width=0.3\textwidth]{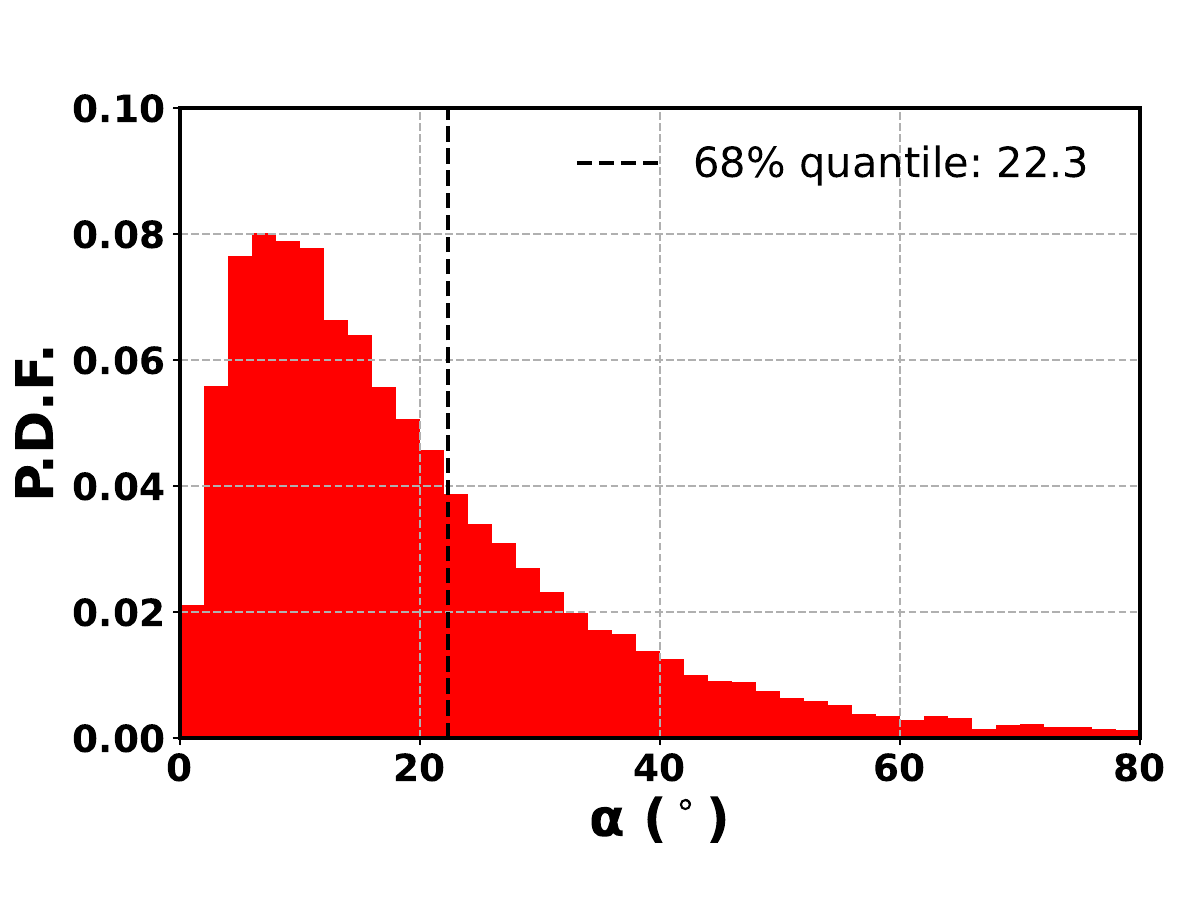}
\includegraphics[width=0.3\textwidth]{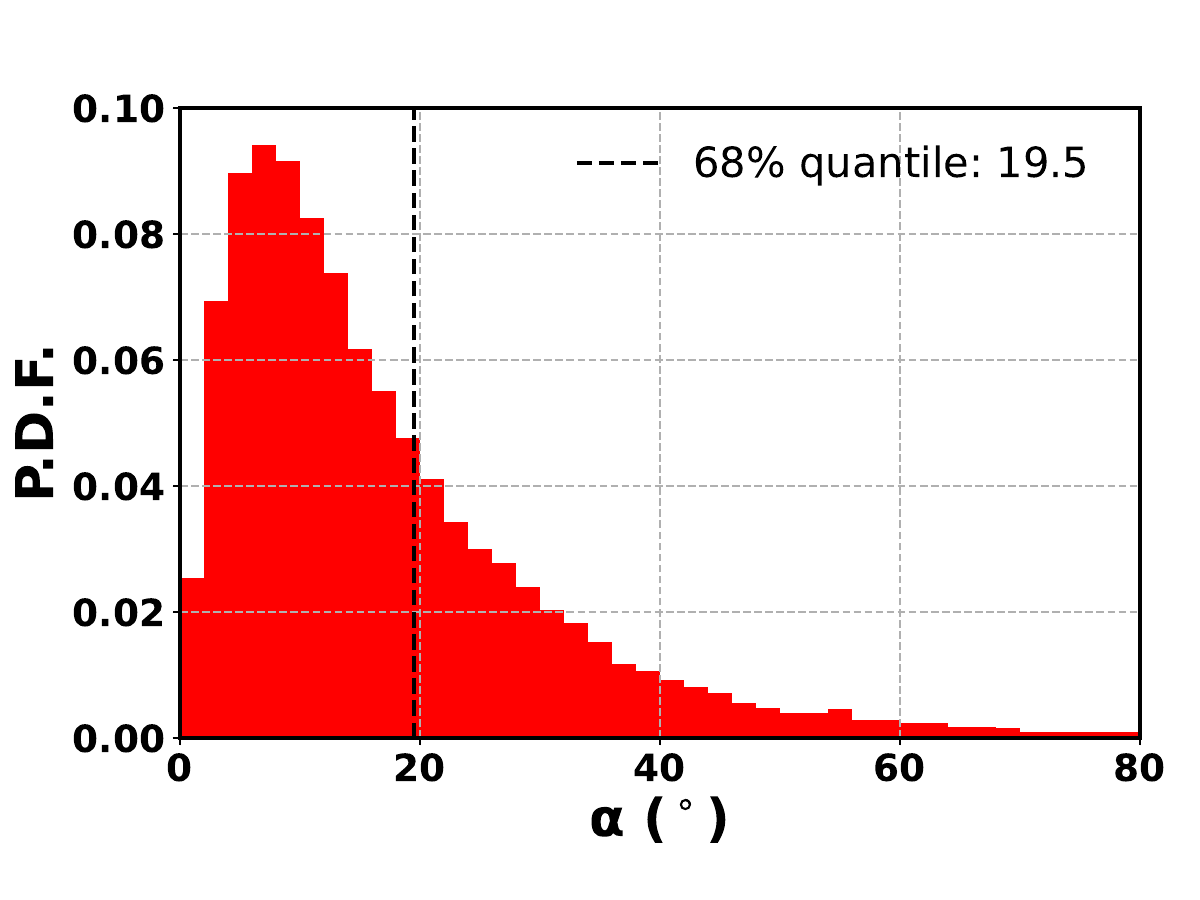}
\includegraphics[width=0.3\textwidth]{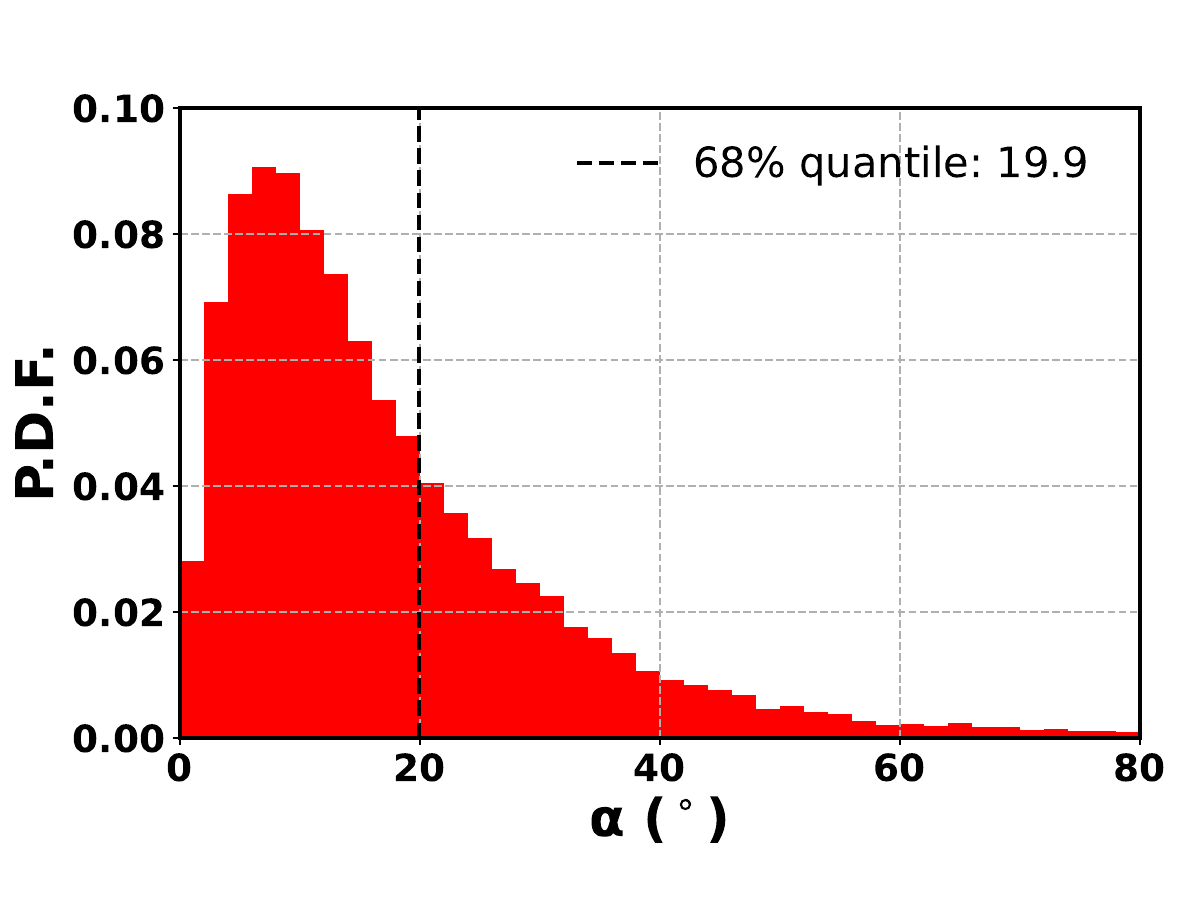}

\includegraphics[width=0.3\textwidth]{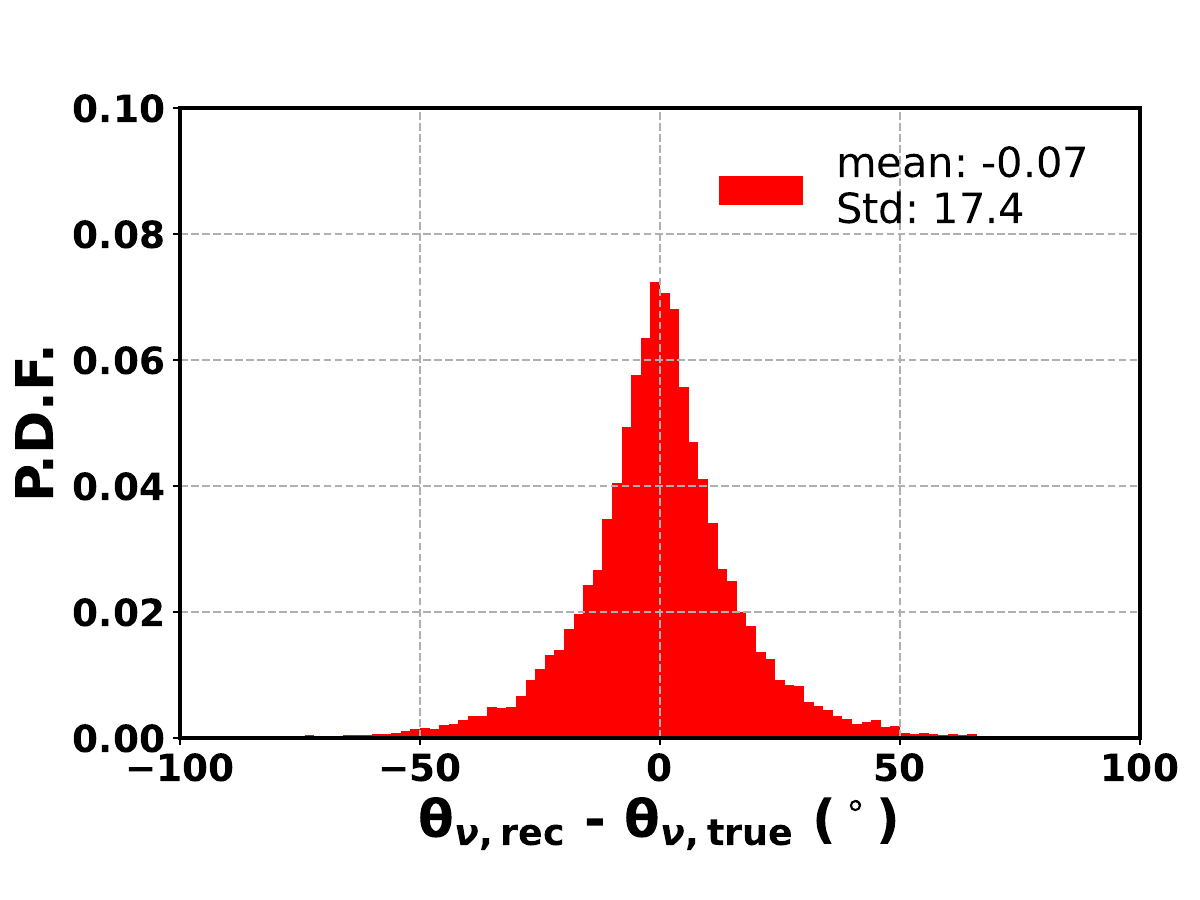}
\includegraphics[width=0.3\textwidth]{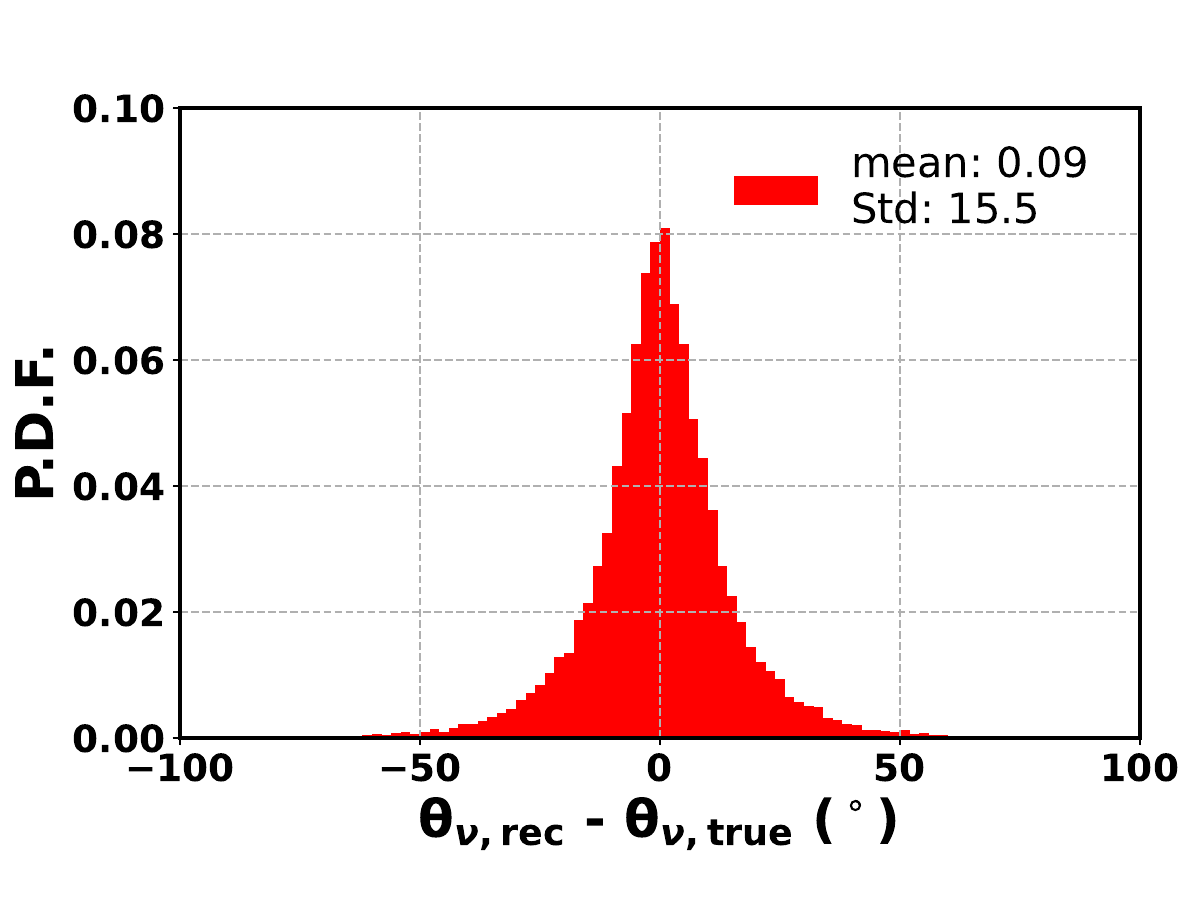}
\includegraphics[width=0.3\textwidth]{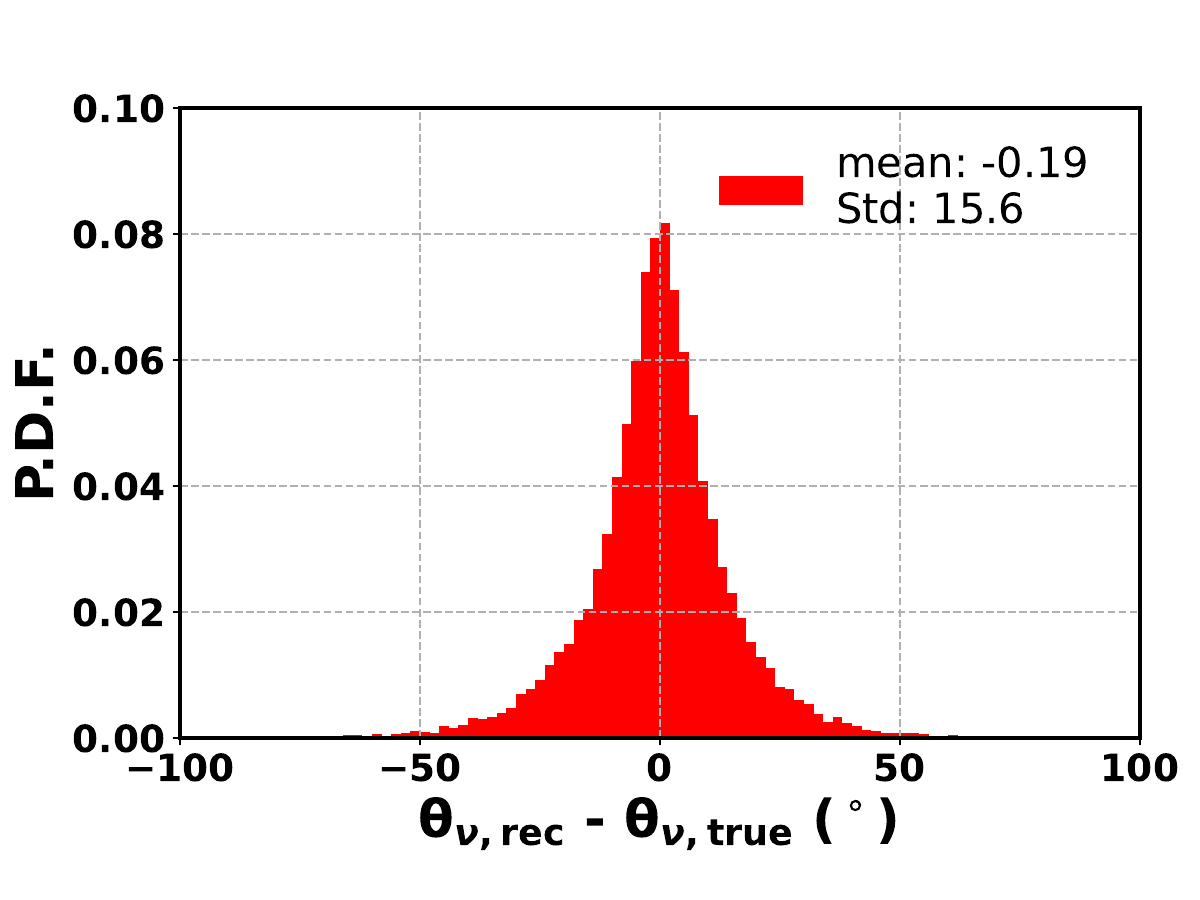}

\subfigure[EfficientNet]{
\hspace{0.5cm}\includegraphics[width=0.3\textwidth]{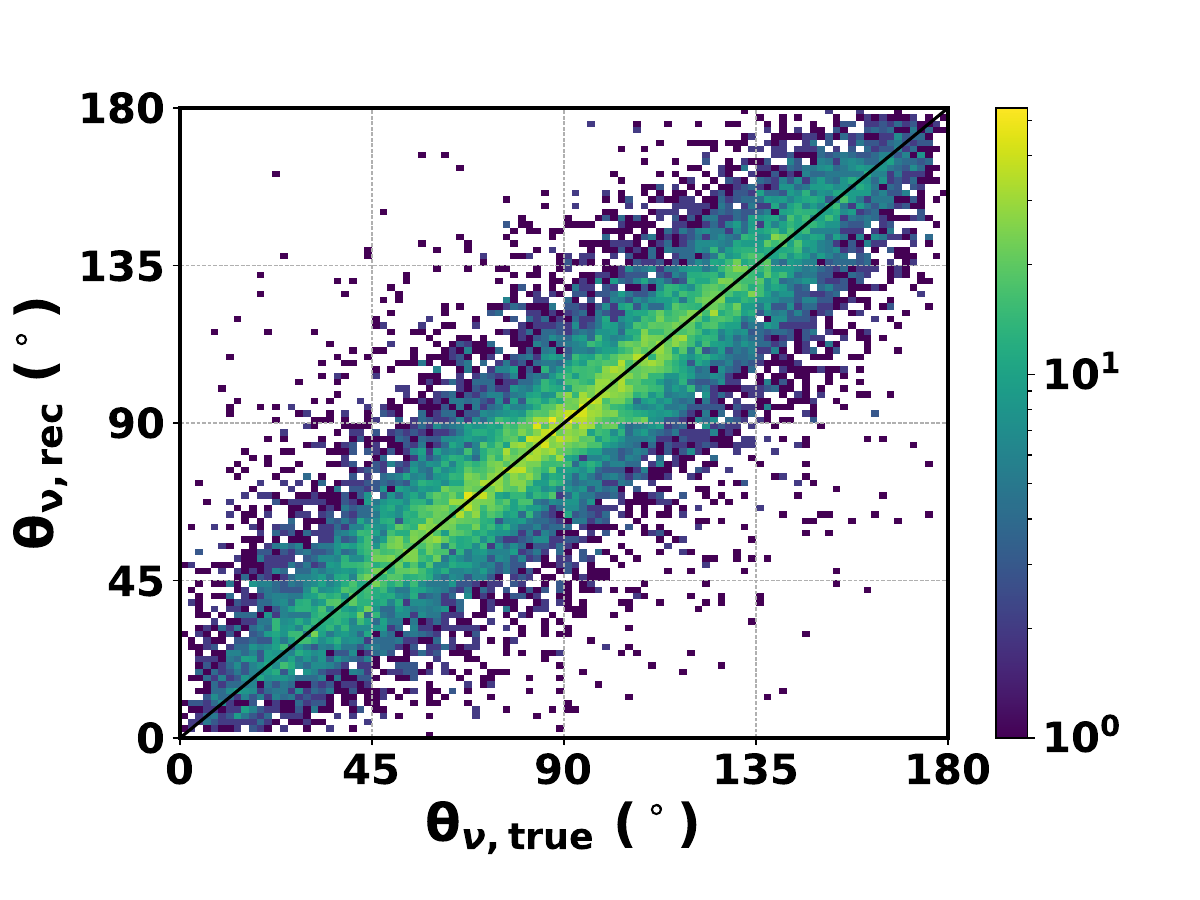}}
\subfigure[DeepSphere]{
\hspace{-0.15cm}\includegraphics[width=0.3\textwidth]{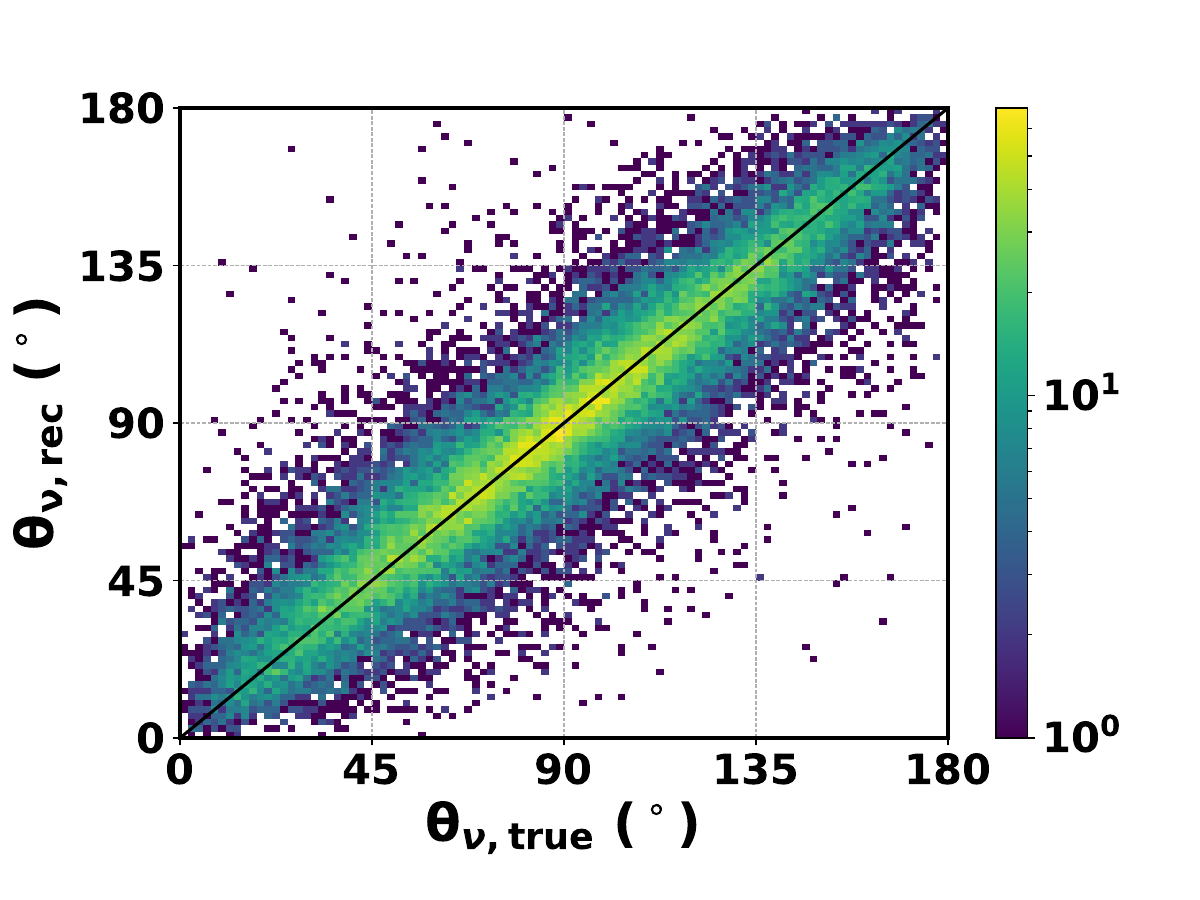}}
\subfigure[PointNet++]{
\hspace{-0.2cm}\includegraphics[width=0.3\textwidth]{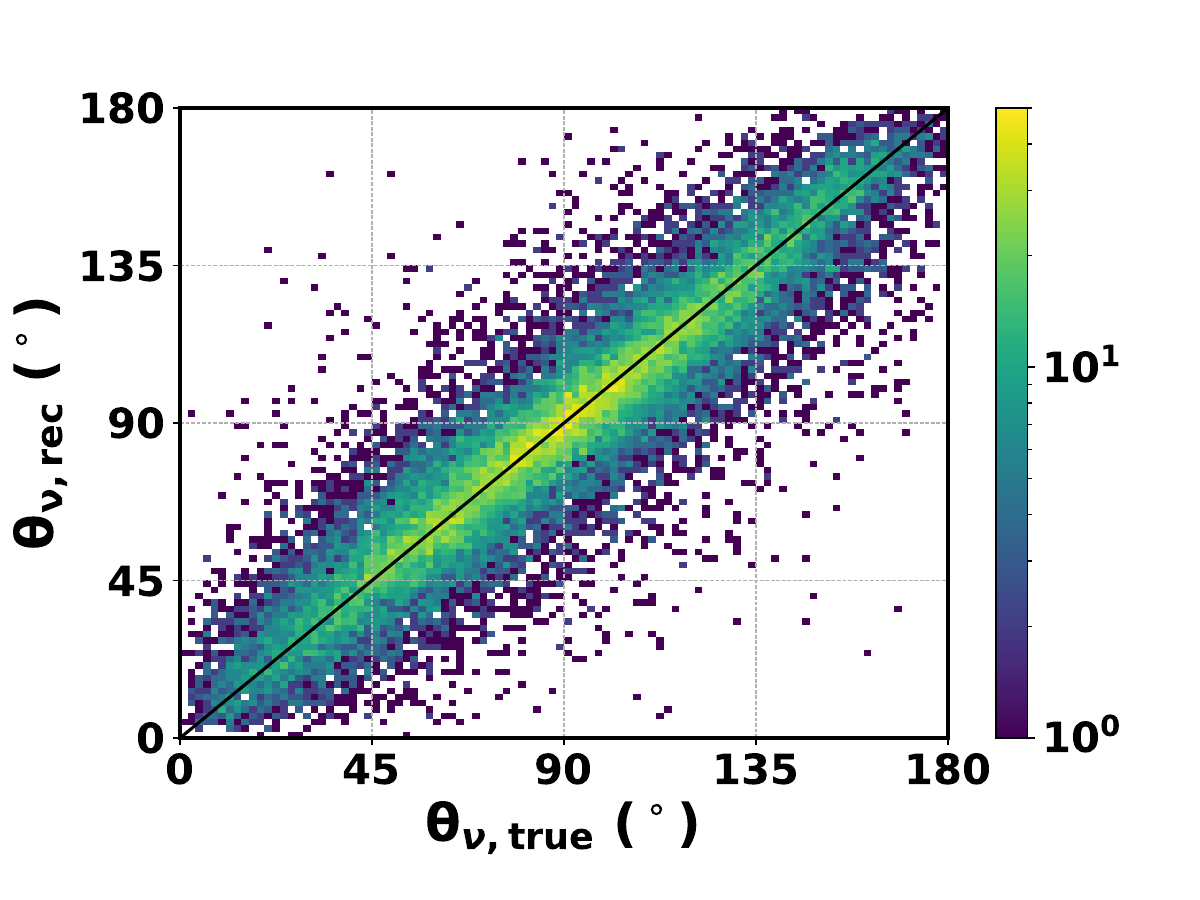}}

\caption{\label{fig:performance_alpha_theta_numu} The directionality reconstruction performance for $\nu_\mu$/$\bar{\nu}_\mu$‐CC events from 1\,GeV to 20\,GeV by (a) EfficientNet, (b) DeepSphere, and (c) PointNet++ models. 
Top: 1D distribution of the opening angle $\alpha$, between the predicted and true neutrino directional vectors. The vertical dashed line marks the 68\% quantile. 
Middle: 1D  distribution of the difference between the predicted and true incoming neutrino zenith angle, $\theta_{\nu, \text{rec}}$ and $\theta_{\nu, \text{true}}$. 
Bottom: Two-dimensional distribution of $\theta_{\nu, \text{rec}}$ vs. $\theta_{\nu, \text{true}}$. The black diagonal line has a slope of 1 for reference.
}
\end{figure*}

\begin{figure*}
\centering
\includegraphics[width=0.3\textwidth]{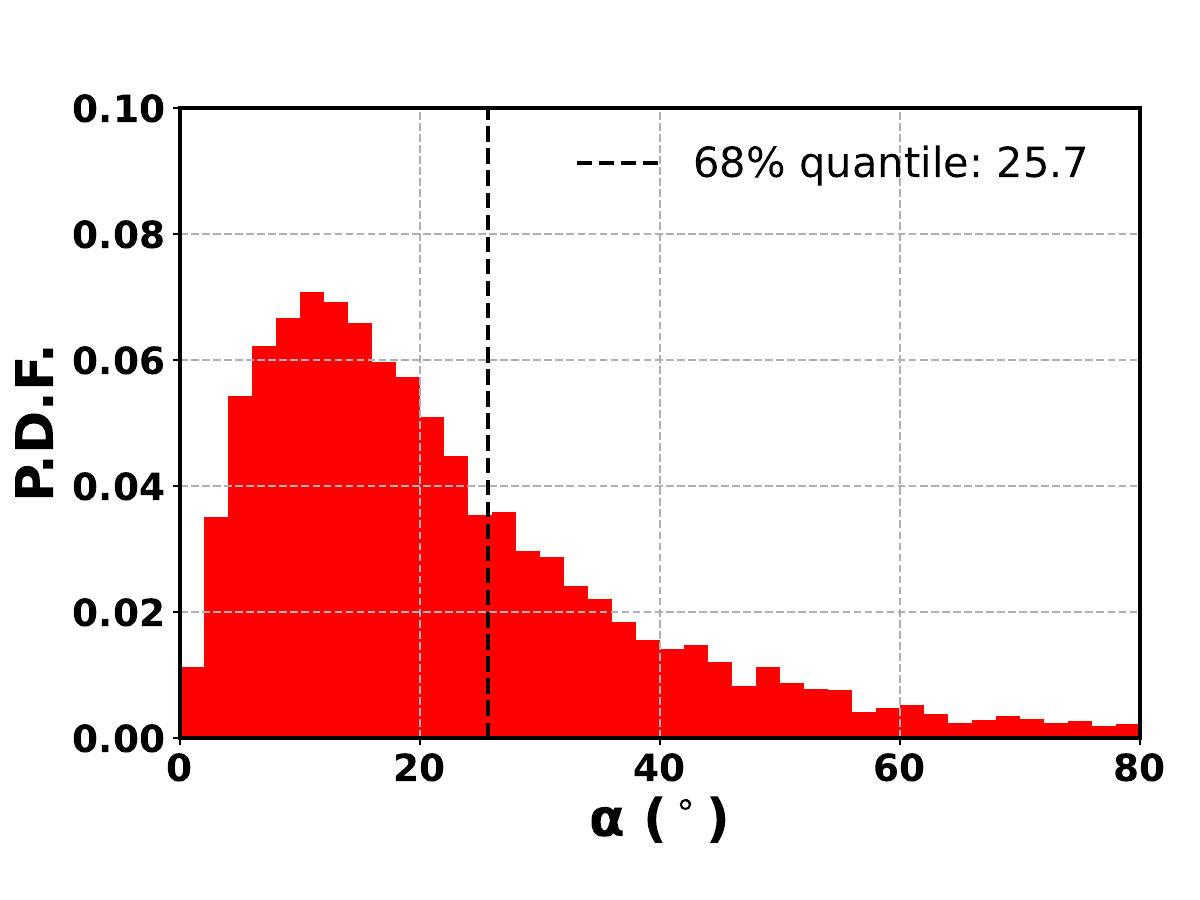}
\includegraphics[width=0.3\textwidth]{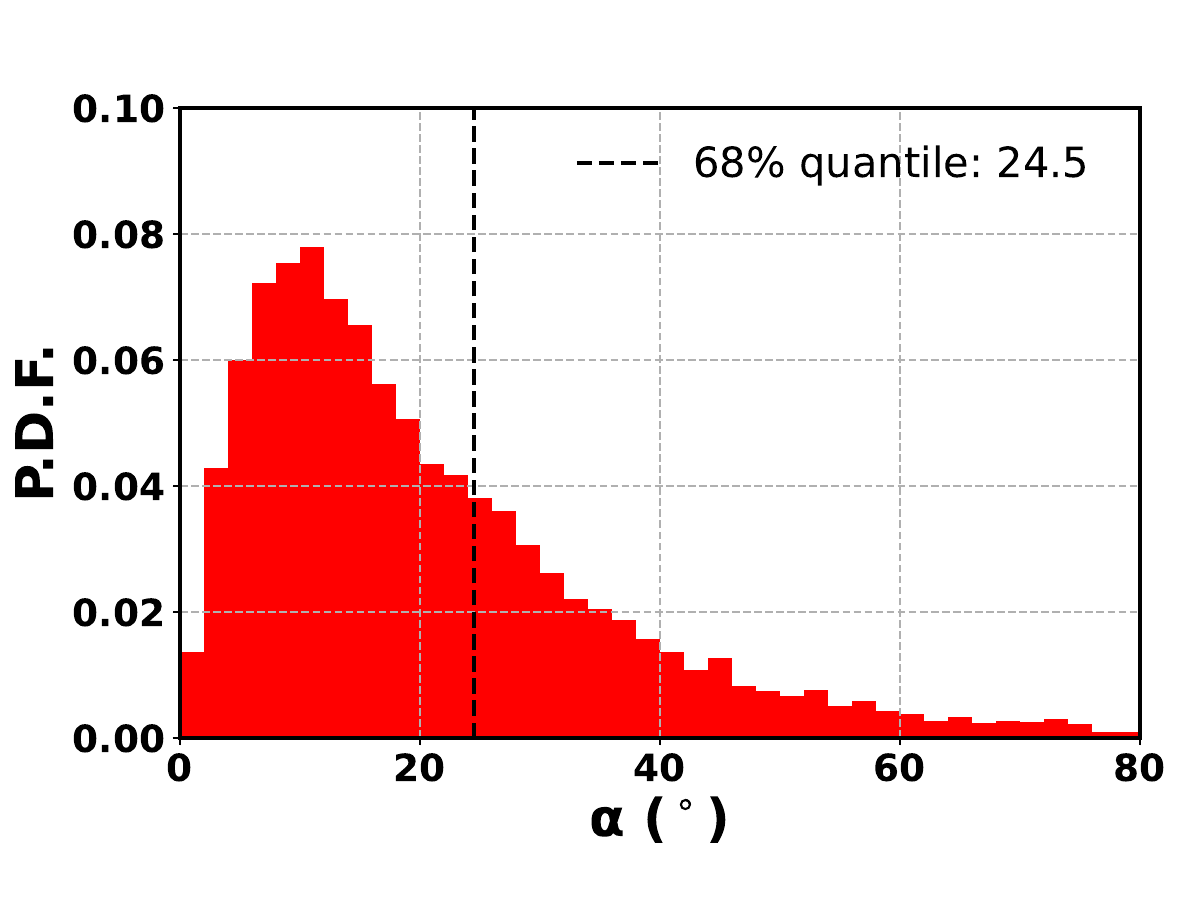}
\includegraphics[width=0.3\textwidth]{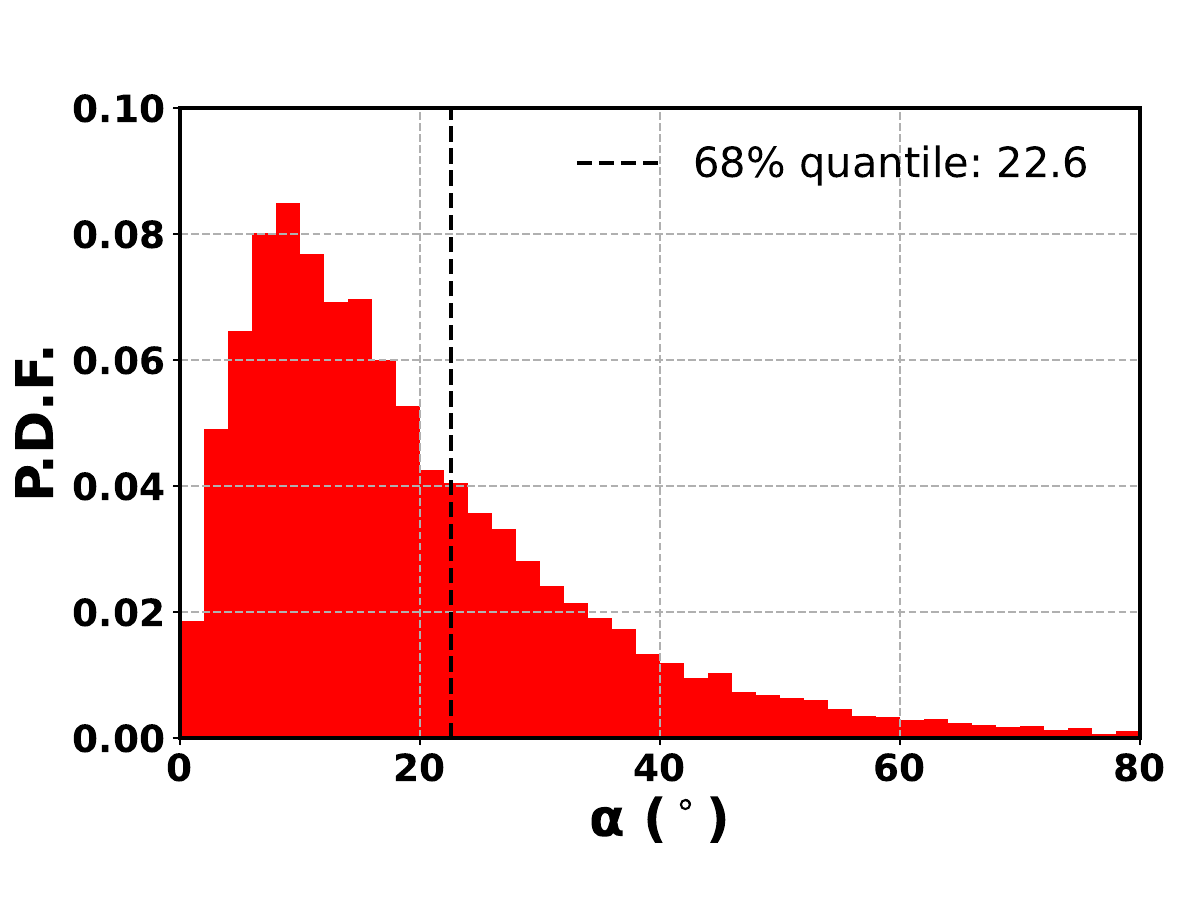}

\includegraphics[width=0.3\textwidth]{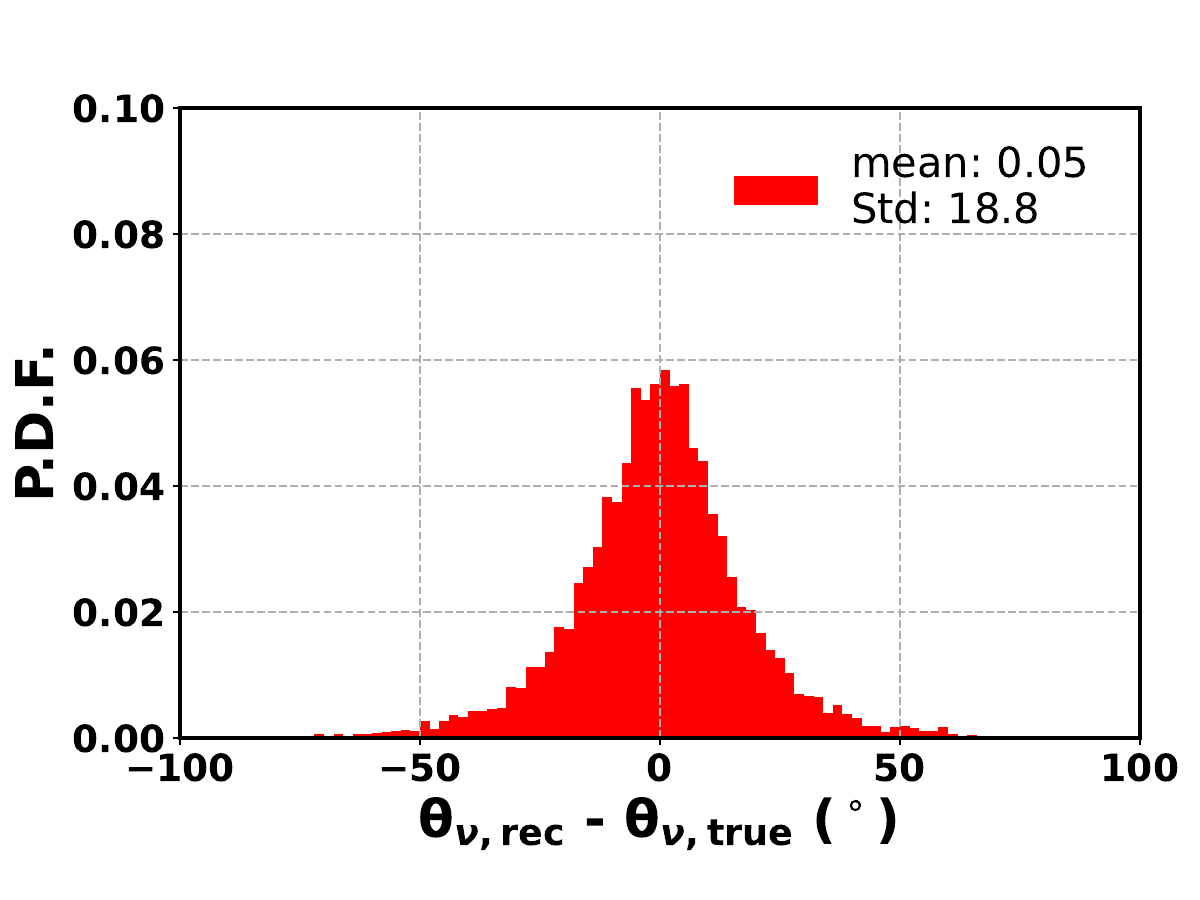}
\includegraphics[width=0.3\textwidth]{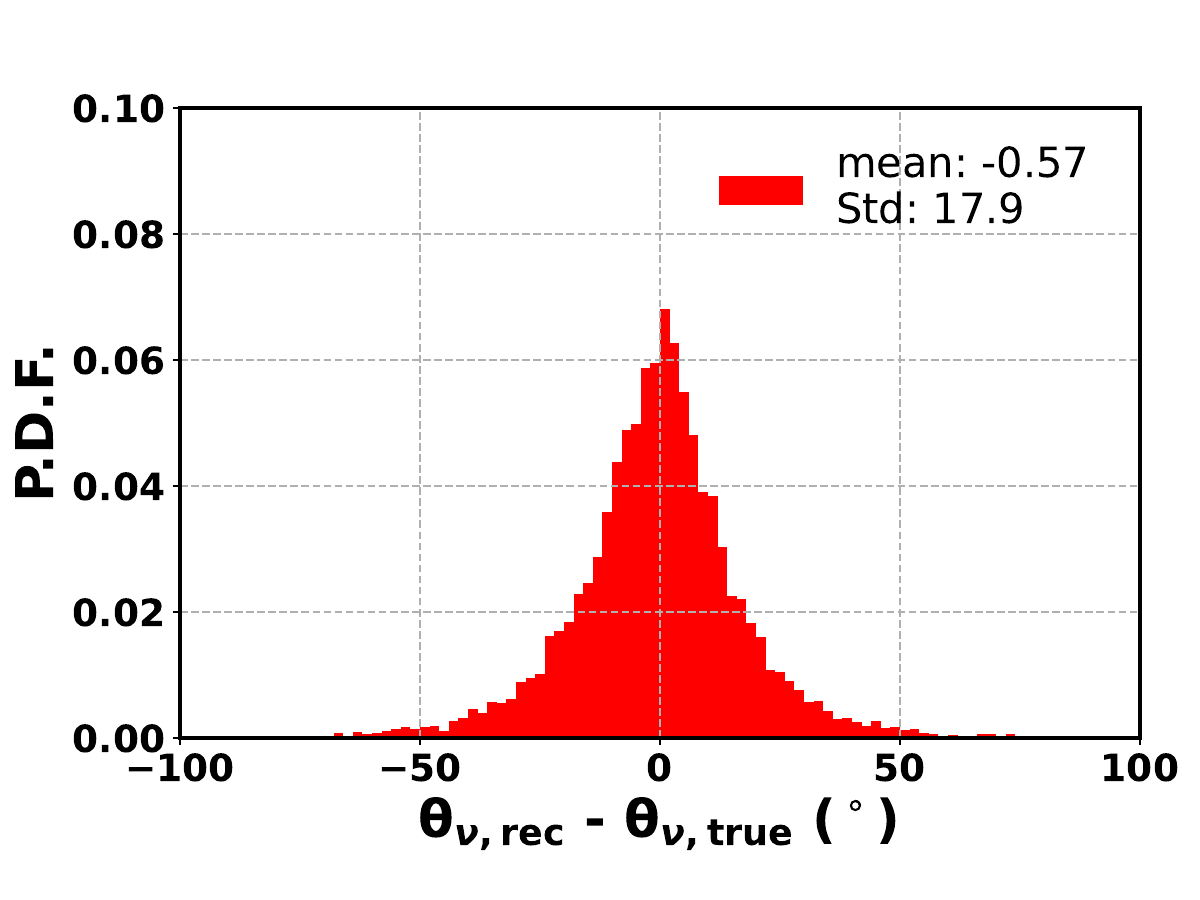}
\includegraphics[width=0.3\textwidth]{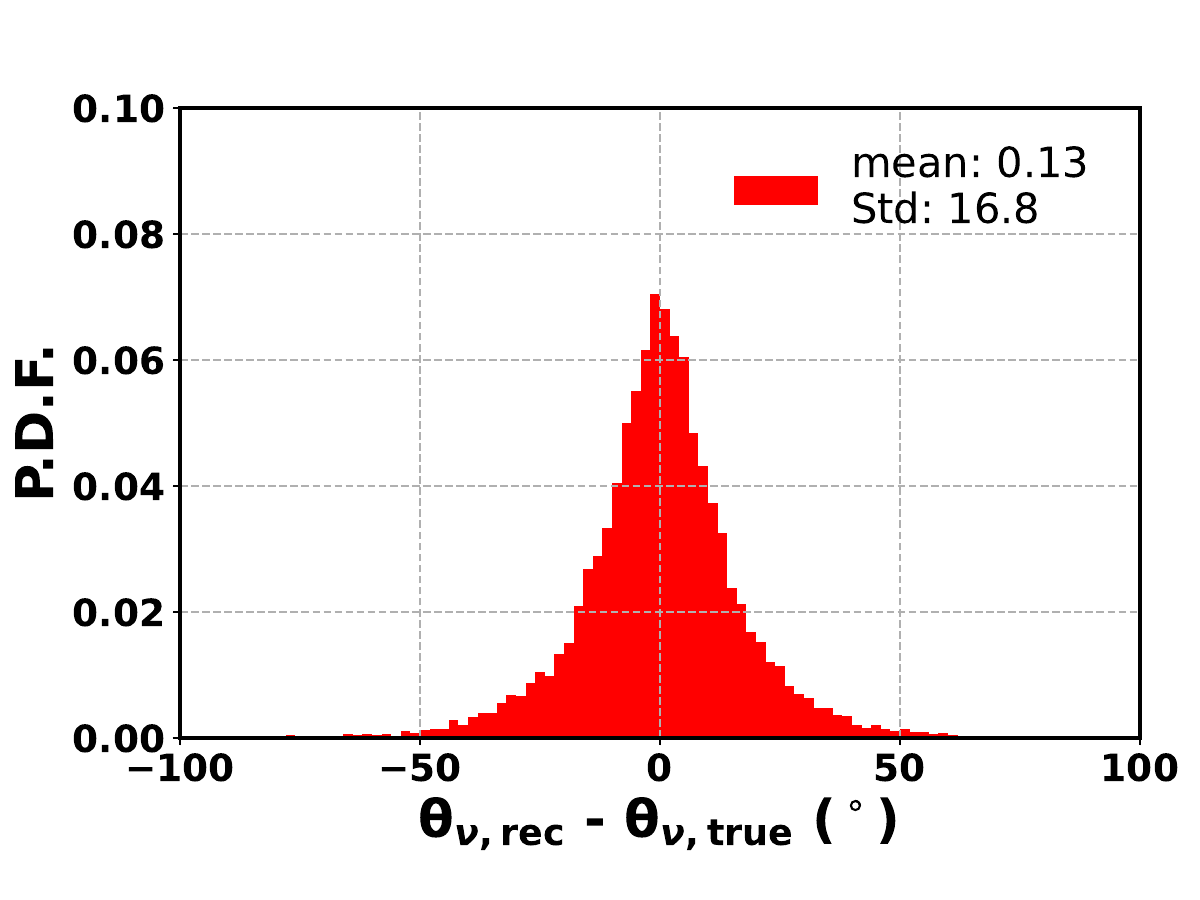}

\subfigure[EfficientNet]{
\hspace{0.5cm}\includegraphics[width=0.3\textwidth]{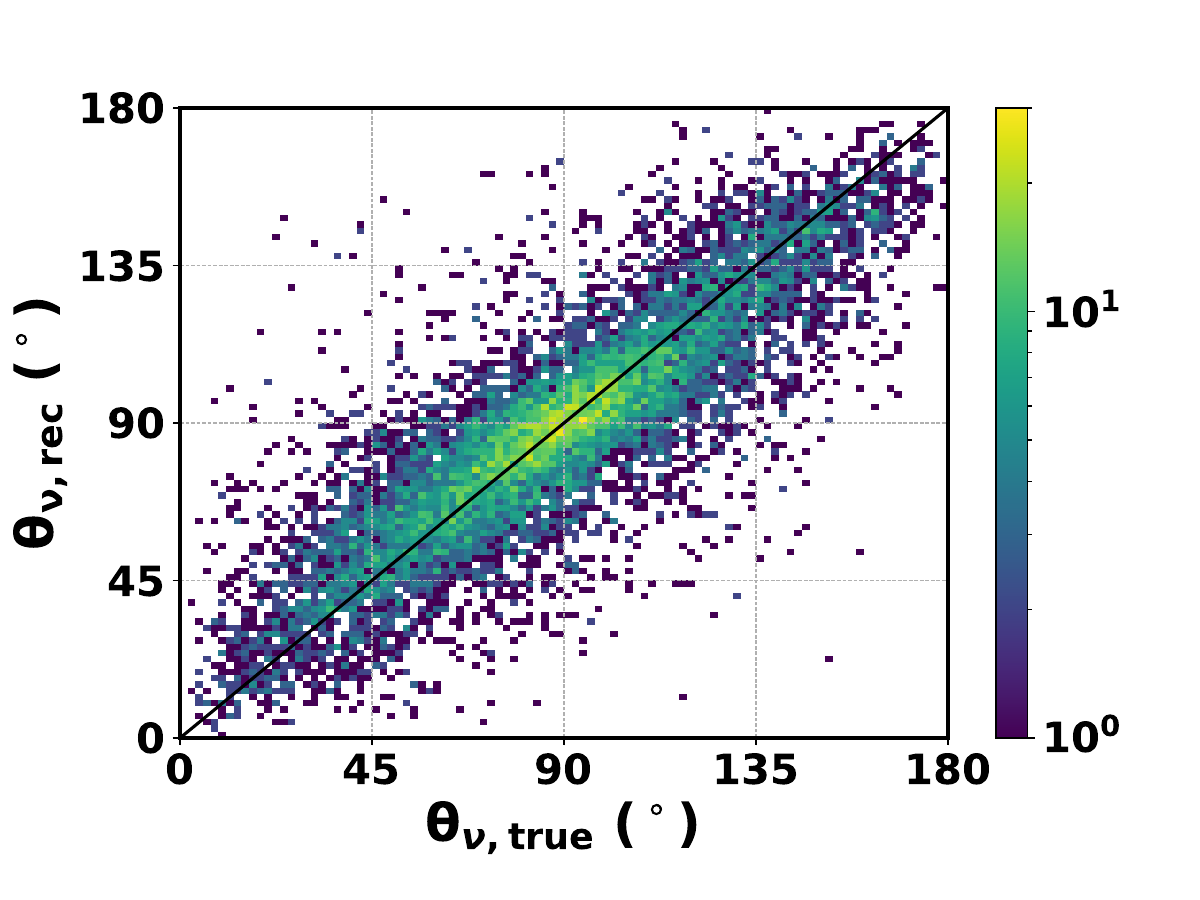}}
\subfigure[DeepSphere]{
\hspace{-0.15cm}\includegraphics[width=0.3\textwidth]{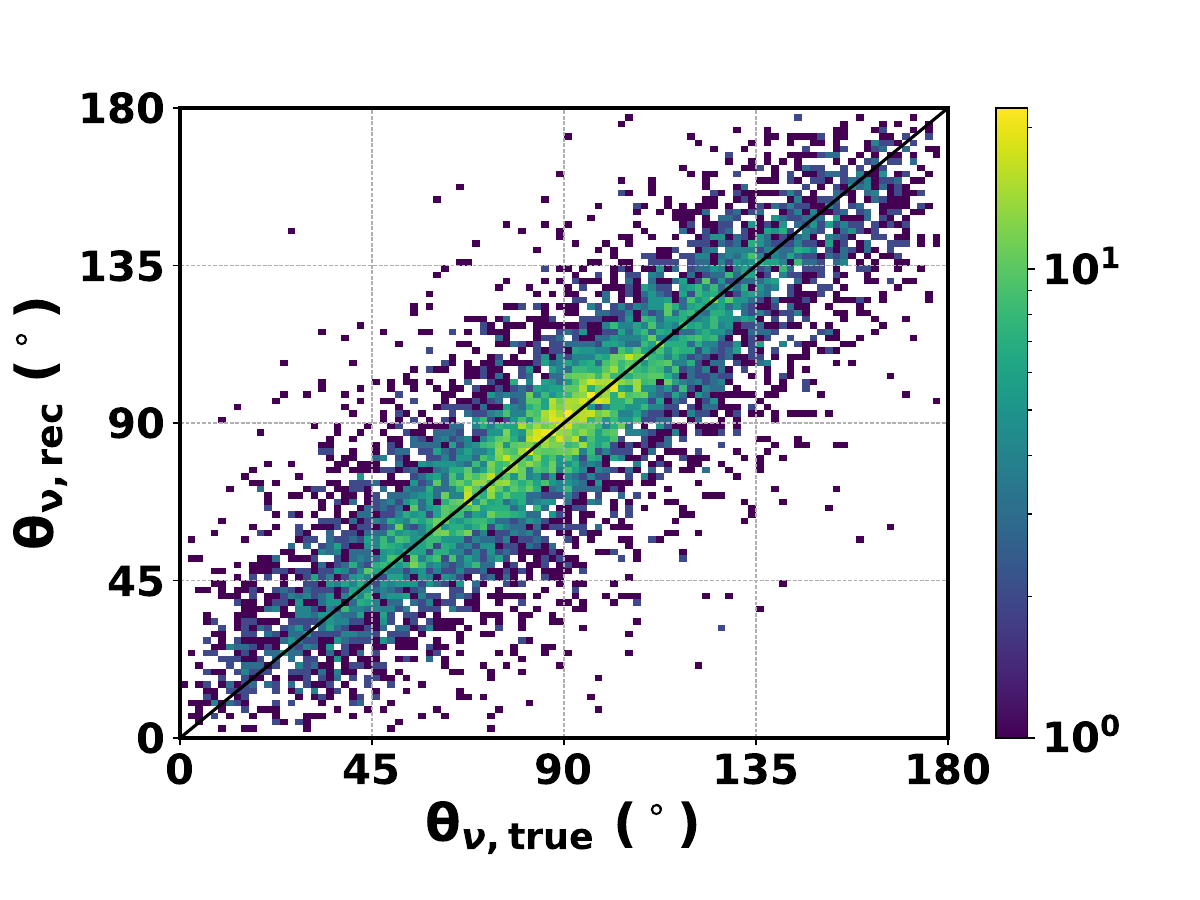}}
\subfigure[PointNet++]{
\hspace{-0.2cm}\includegraphics[width=0.3\textwidth]{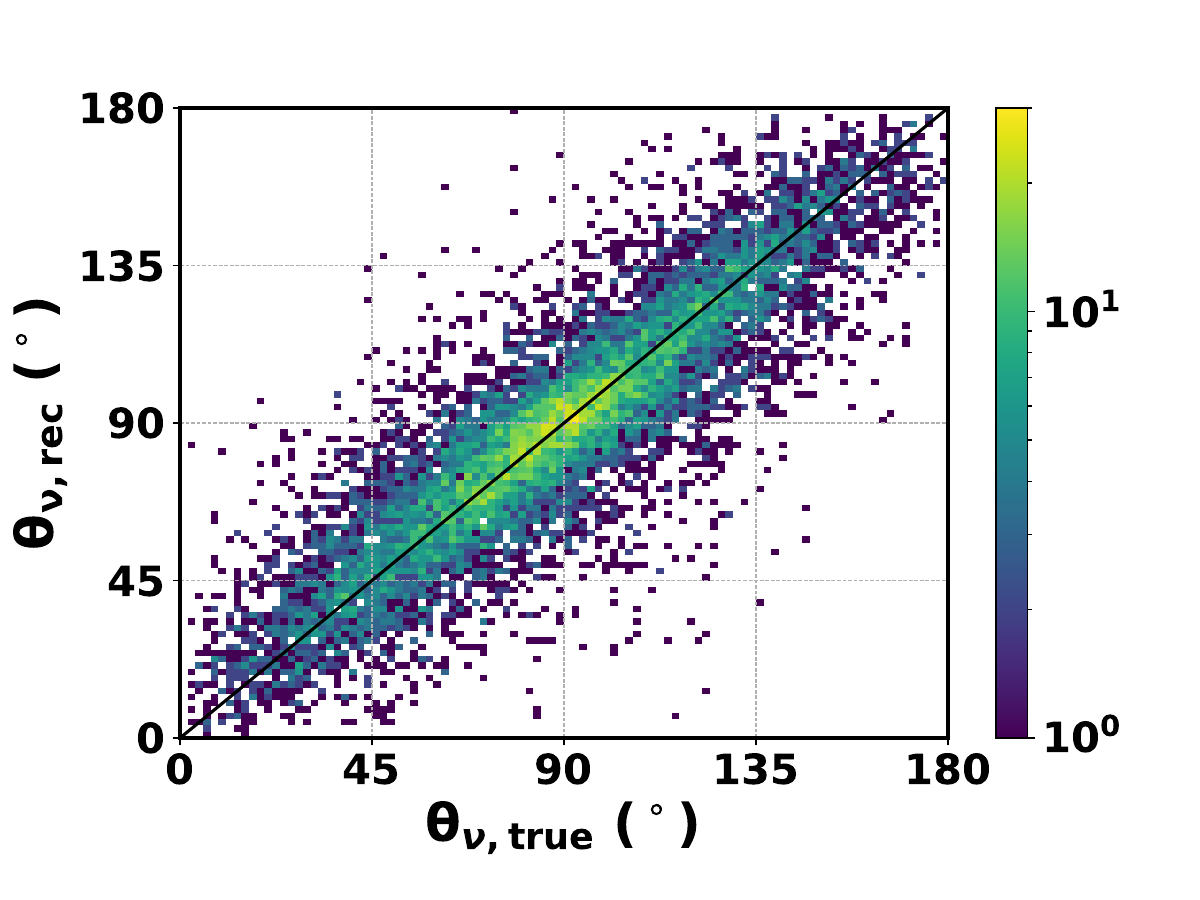}}
\caption{\label{fig:performance_alpha_theta_nue} The directionality reconstruction performance for $\nu_e$/$\bar{\nu}_e$‐CC events from 1\,GeV to 20\,GeV by (a) EfficientNet, (b) DeepSphere, and (c) PointNet++ models. 
Top: 1D distribution of the opening angle, $\alpha$, between the predicted and true neutrino directional vectors. The vertical dashed line marks the 68\% quantile. 
Middle: 1D  distribution of the difference between the predicted and true incoming neutrino zenith angle, $\theta_{\nu, \text{rec}}$ and $\theta_{\nu, \text{true}}$. 
Bottom: Two-dimensional distribution of $\theta_{\nu, \text{rec}}$ vs. $\theta_{\nu, \text{true}}$. The black diagonal line has a slope of 1 for reference.
}
\end{figure*}

For each model mentioned above, 80\% of the $\nu_\mu$/$\bar{\nu}_\mu$ and $\nu_e$/$\bar{\nu}_e$‐CC samples described in section \ref{sec:sim} are used for training separately, with the other 20\% used for validations.
The results of the validation sample are quoted as the reconstruction performances.  

The performances are evaluated in two ways: firstly by evaluating the opening angle $\alpha$ between the true and reconstructed neutrino directions,
and secondly by evaluating the difference 
between the true and reconstructed zenith angle of the incoming neutrino ($\theta_\nu$).
Both $\alpha$ and $\theta_\nu$ are illustrated in Fig.~\ref{fig:angles}.
The evaluation is done in 1\,GeV neutrino energy bins since the performances can be energy-dependent, and also for the whole sample from 1\,GeV to 20\,GeV. 
The zenith angle $\theta_\nu$ is of particular interest since $\cos{\theta_\nu}$ is the direct input to atmospheric neutrino oscillation measurements. 
The resolution on $\alpha$ ($\sigma_\alpha$) is defined as the 68\% quantile of the $\alpha$ distribution, given that $\alpha$ is always larger than 0.
 For $\theta_\nu$, the resolution in each energy bin ($\sigma_{\theta_\nu}$) is defined as 
the standard deviation of the Gaussian fit to the distribution of differences between the reconstructed and true values ($\theta_{\nu,\text{rec}} - \theta_{\nu,\text{true}}$), since these distributions are found to be approximately Gaussian.
While for the entire sample, the standard deviation of the $\theta_{\nu,\text{rec}} - \theta_{\nu,\text{true}}$ distribution (Std$_{\theta\nu}$) is used in place of the Gaussian $\sigma_{\theta_\nu}$. This substitution is made due to the non-Gaussian shape resulting from the superposition of all the energies.

The resulting performances are shown in Fig.~\ref{fig:performance_alpha_theta_numu} and Fig.~\ref{fig:performance_alpha_theta_nue} for $\nu_\mu$/$\bar{\nu}_\mu$ and $\nu_e$/$\bar{\nu}_e$‐CC events from 1\,GeV to 20\,GeV respectively. 
The plots on the top row show the distributions of $\alpha$, the distributions of $\theta_{\nu,\text{rec}} - \theta_{\nu,\text{true}}$ are shown in the middle row, and the $\theta_{\nu,\text{rec}}$  vs $\theta_{\nu,\text{true}}$ 2D plots are shown in the bottom row. From left to right the three columns correspond to the three models: EfficientNet, DeepSphere and PointNet++ respectively. 
The reconstruction results are also summarized in Tab.~\ref{table:performance_table}.
Overall the three models show comparable performances, with the maximum difference between models around 2$^{\circ}$. 
No obvious reconstruction bias is observed.

\begin{table}
\centering
\caption{Summary of the reconstruction performances for $\nu_\mu$/$\bar{\nu}_\mu$‐CC and $\nu_e$/$\bar{\nu}_e$‐CC events using three different models. 
The performances on $\alpha$ ($\sigma_{\alpha}$) are benchmarked by the 68\% quantile of the $\alpha$ distribution. 
The performances on  $\theta_\nu$ are evaluated by the standard deviation values of the $\theta_{\nu,\text{rec}} - \theta_{\nu,\text{true}}$ distributions given their non-Gaussian shapes. 
}
\begin{tabular}{lcccc}

 \hline
 \hline
 &  & EfficientNet-V2 & DeepSphere & PointNet++ \\
 \hline \multirow{2}{2em}{$\nu_\mu$/$\bar{\nu}_\mu$} 
 	& $\sigma_{\alpha}$ & $22.3^{\circ}$  & $19.5^{\circ}$  & $19.9^{\circ}$ \\
        & Std$_{\theta_{\nu}}$  & $17.4^{\circ}$  & $15.5^{\circ}$  & $15.6^{\circ}$\\ 
 \hline \multirow{2}{2em}{$\nu_e$/$\bar{\nu}_e$} 
 	& $\sigma_{\alpha}$  & $25.7^{\circ}$  & $24.5^{\circ}$  & $22.6^{\circ}$ \\
        & Std$_{\theta_{\nu}}$  & $18.8^{\circ}$  & $17.9^{\circ}$  & $16.8^{\circ}$\\ 
 \hline 
  \hline 
\end{tabular}
\label{table:performance_table}
\end{table}

\begin{figure*}
\centering
\includegraphics[width=0.45\textwidth]{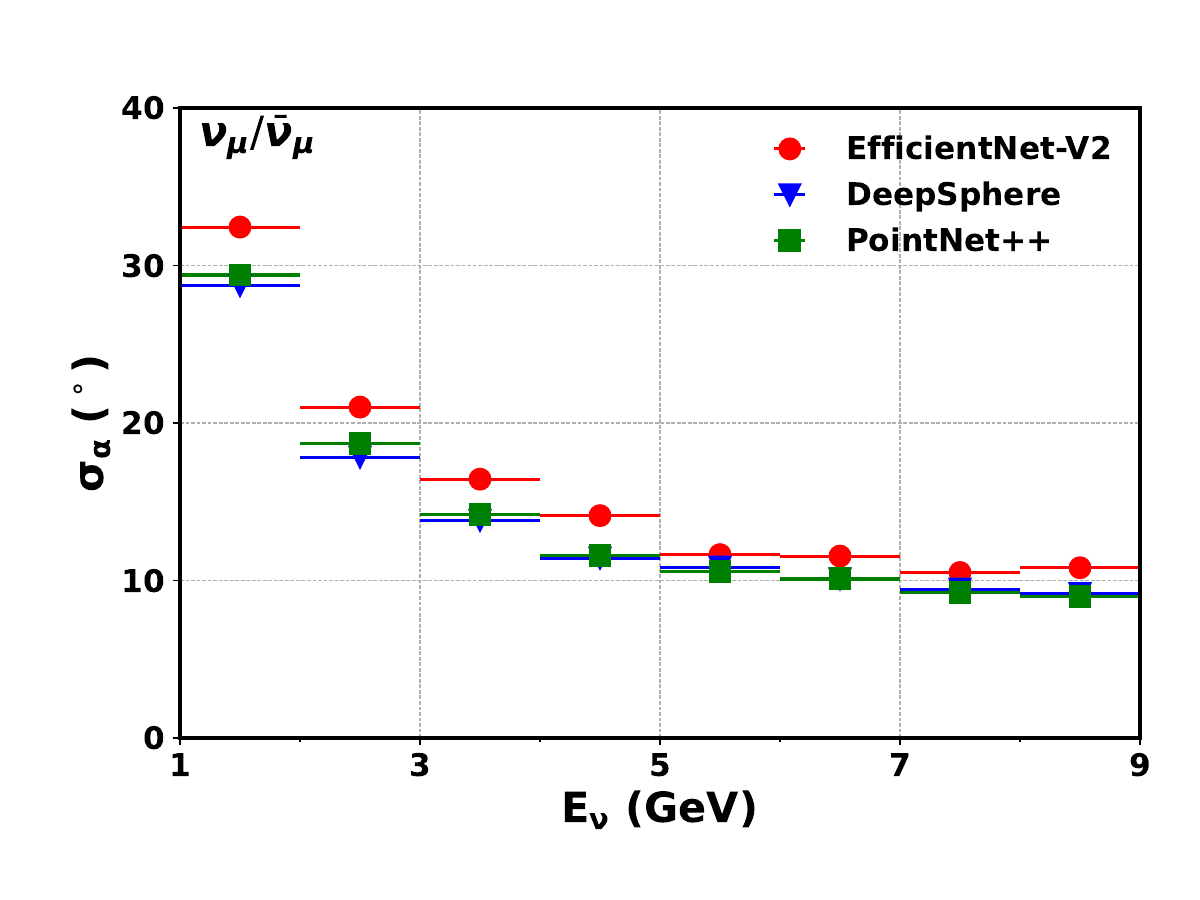}
\includegraphics[width=0.45\textwidth]{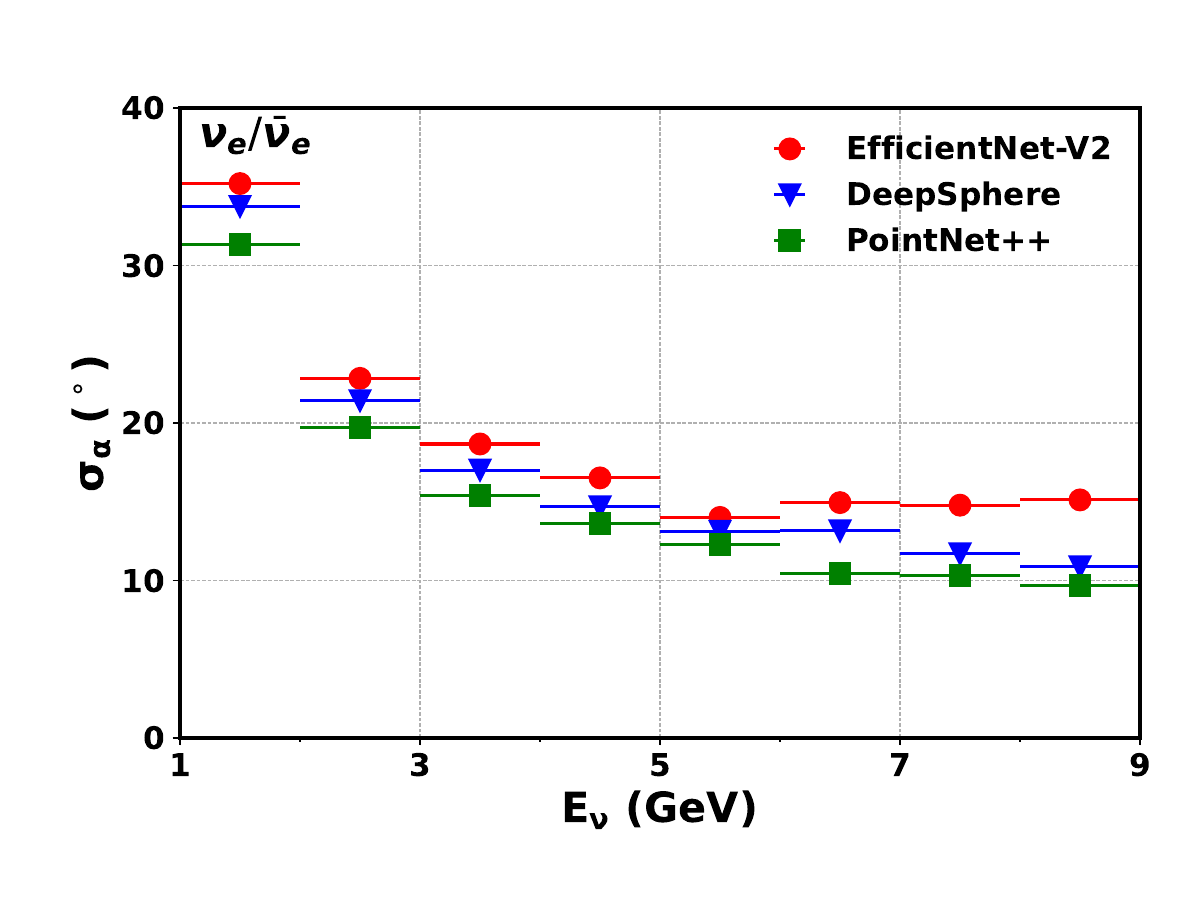}
\subfigure[$\nu_\mu$/$\bar{\nu}_\mu$‐CC]{\hspace{-1.45cm}
\hspace{1.5cm}\includegraphics[width=0.45\textwidth]{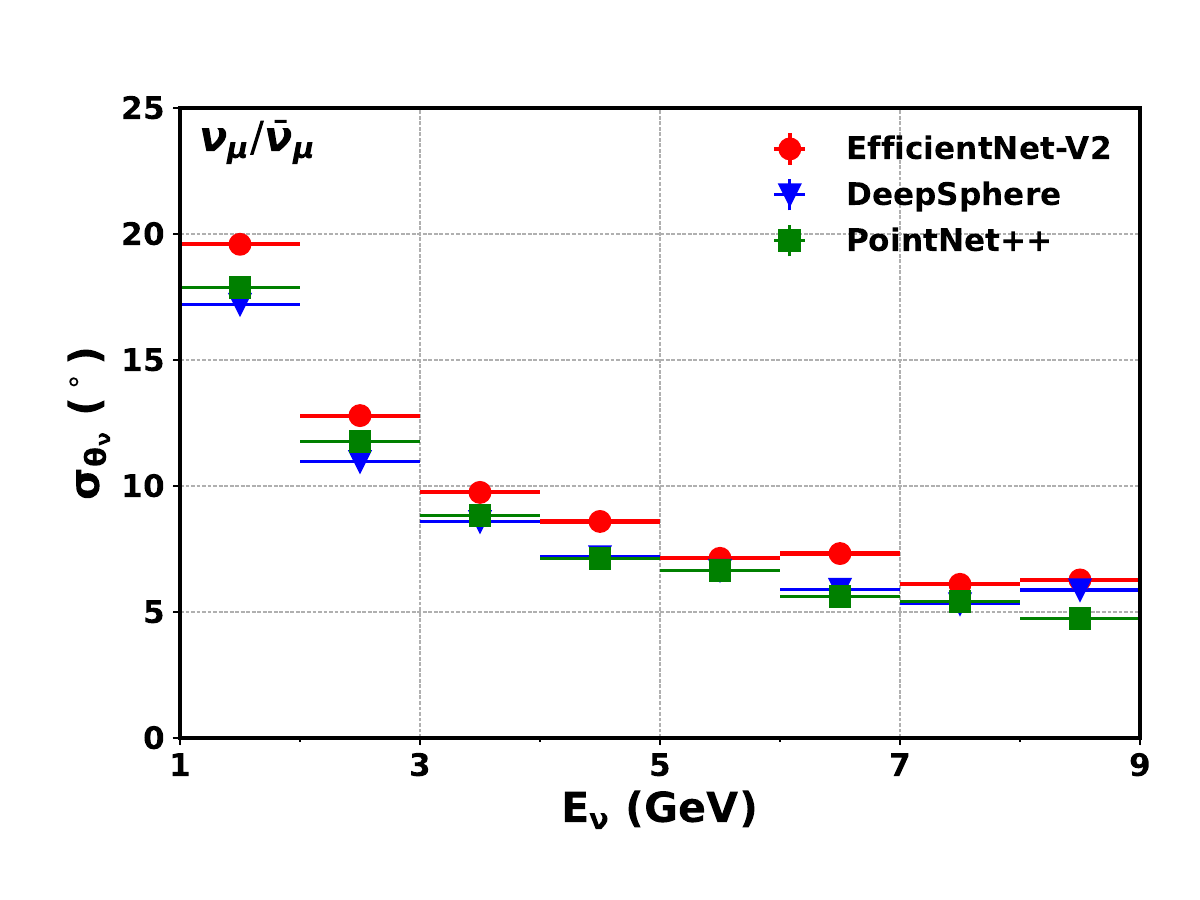}}
\subfigure[$\nu_e$/$\bar{\nu}_e$‐CC]{\hspace{-0.5cm}
\hspace{0.3cm}\includegraphics[width=0.45\textwidth]{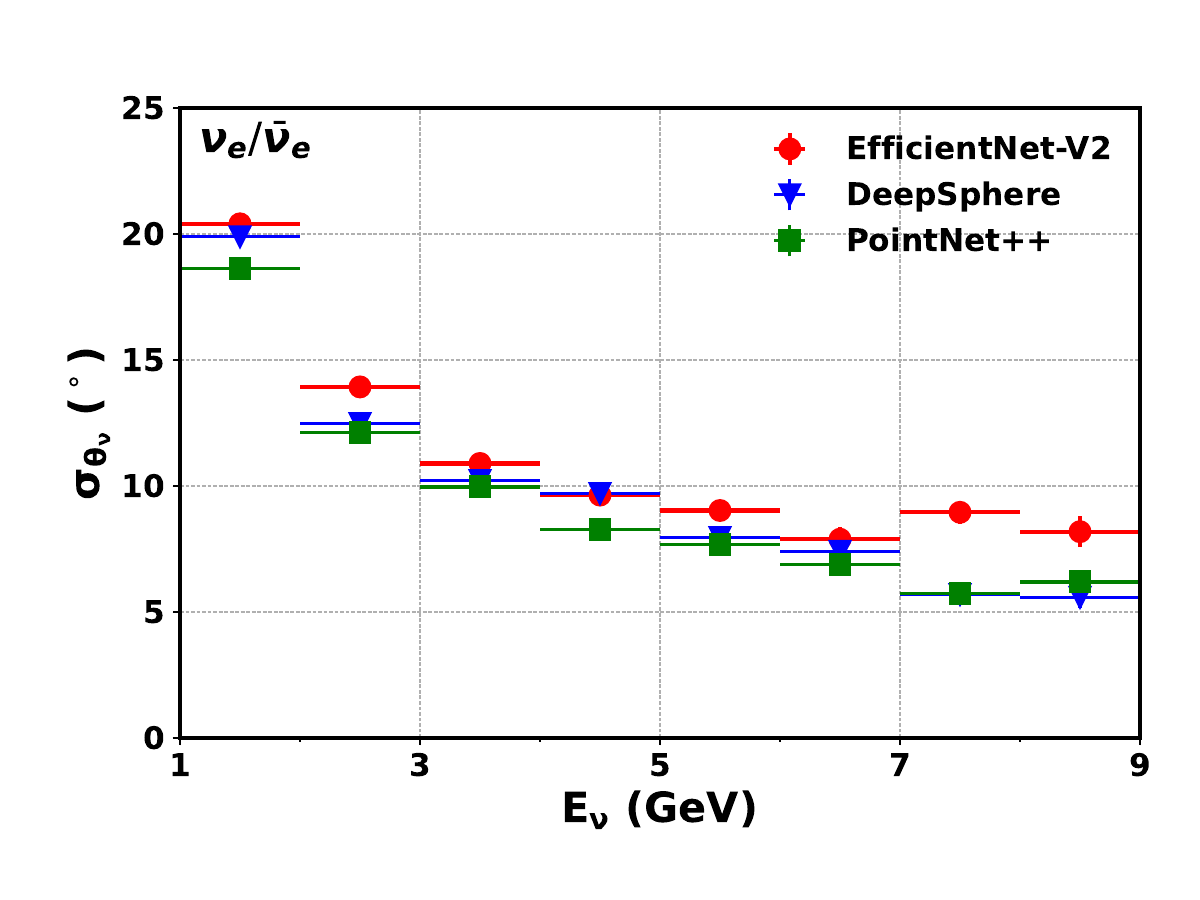}}
\caption{\label{fig:sigma_alpha_theta} 
The $\alpha$ (top) and $\theta_\nu$ (bottom) resolutions are shown as a function of neutrino energy $E_\nu$ for (a) $\nu_\mu$/$\bar{\nu}_\mu$‐CC and (b) $\nu_e$/$\bar{\nu}_e$‐CC events in the three models. The resolution improves with increasing $E_\nu$. The $\nu_\mu$/$\bar{\nu}_\mu$‐CC events in general have better resolution than the $\nu_e$/$\bar{\nu}_e$‐CC events at the same energy.
}
\end{figure*}

The $\alpha$ and $\theta_\nu$ resolutions as functions of the incoming neutrino energy are shown in Fig.~\ref{fig:sigma_alpha_theta}.
As one would expect the performance gets better as the energy increases for both neutrino flavors, and a consistent trend is observed for the three different models. 
Moreover, for neutrinos with the same energy, $\nu_\mu$/$\bar{\nu}_\mu$‐CC has better resolution than $\nu_e$/$\bar{\nu}_e$‐CC.
This is also expected since the muon track in the final state of $\nu_\mu$/$\bar{\nu}_\mu$‐CC interactions generally exhibits stronger directionality than the electron shower in the case of $\nu_e$/$\bar{\nu}_e$‐CC at the same energy.

\section{Discussion \label{sec:dis}}
\begin{figure*}
\centering
\includegraphics[width=0.45\textwidth]{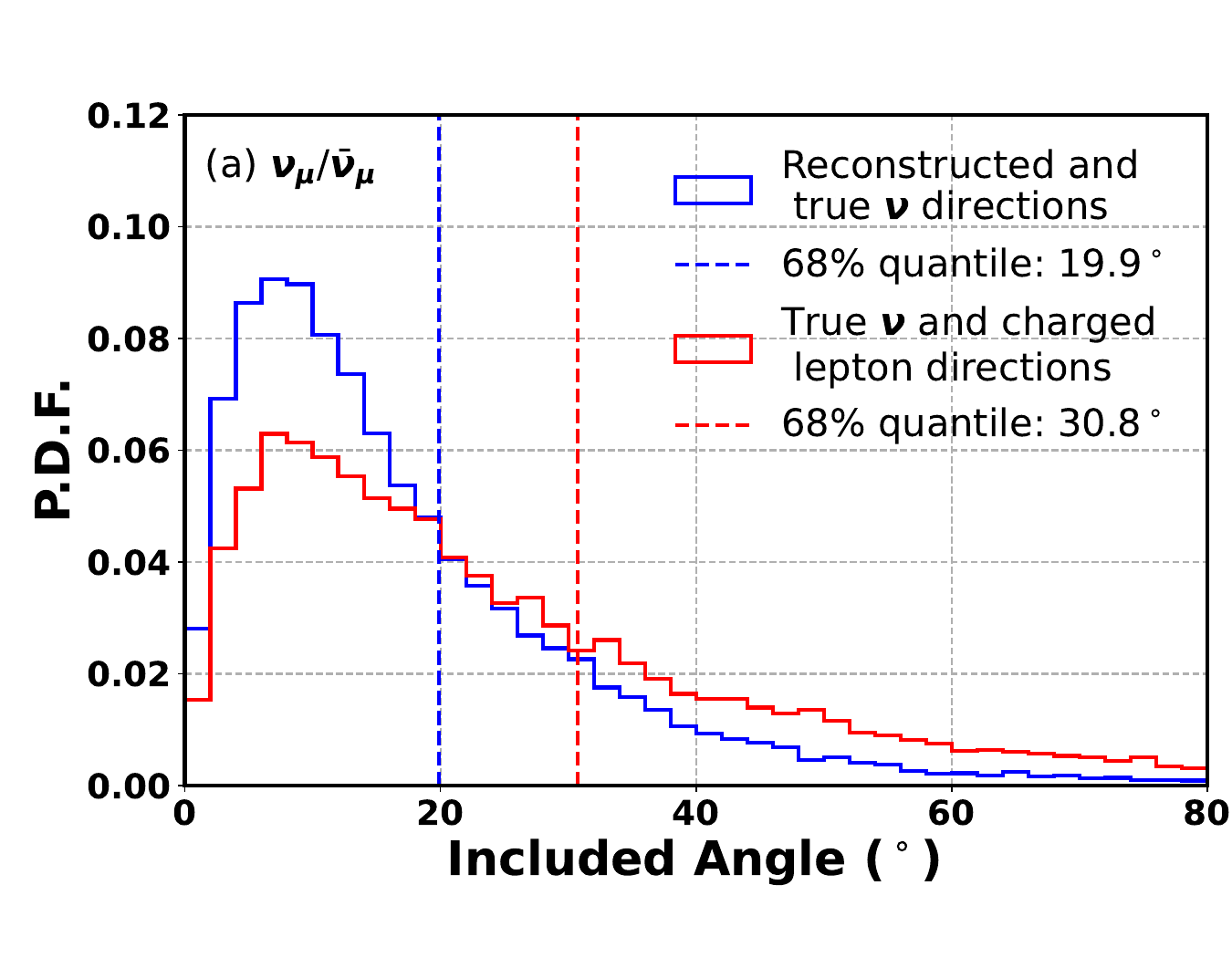}
\includegraphics[width=0.45\textwidth]{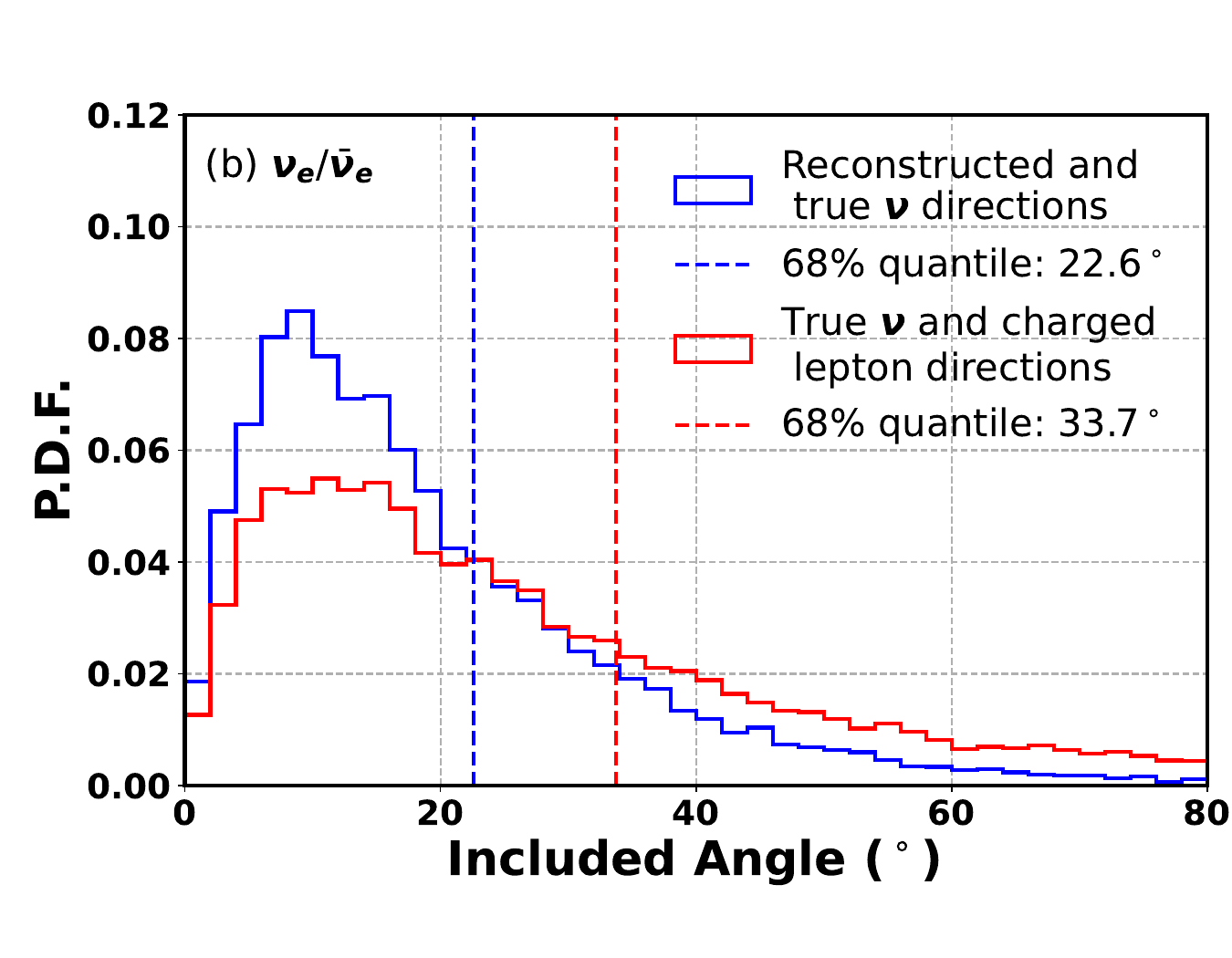}
\caption{\label{fig:included_angles} Comparison between two included angles: the one between the true and reconstructed neutrino direction from PointNet++ in this study (blue lines), and the one between the incident neutrino and final-state charged lepton directions (red lines) using the same (a) $\nu_\mu$/$\bar{\nu}_\mu$‐CC and (b) $\nu_e$/$\bar{\nu}_e$‐CC samples. 
}
 
\end{figure*}

\begin{figure*}
\centering
\includegraphics[width=0.45\textwidth]{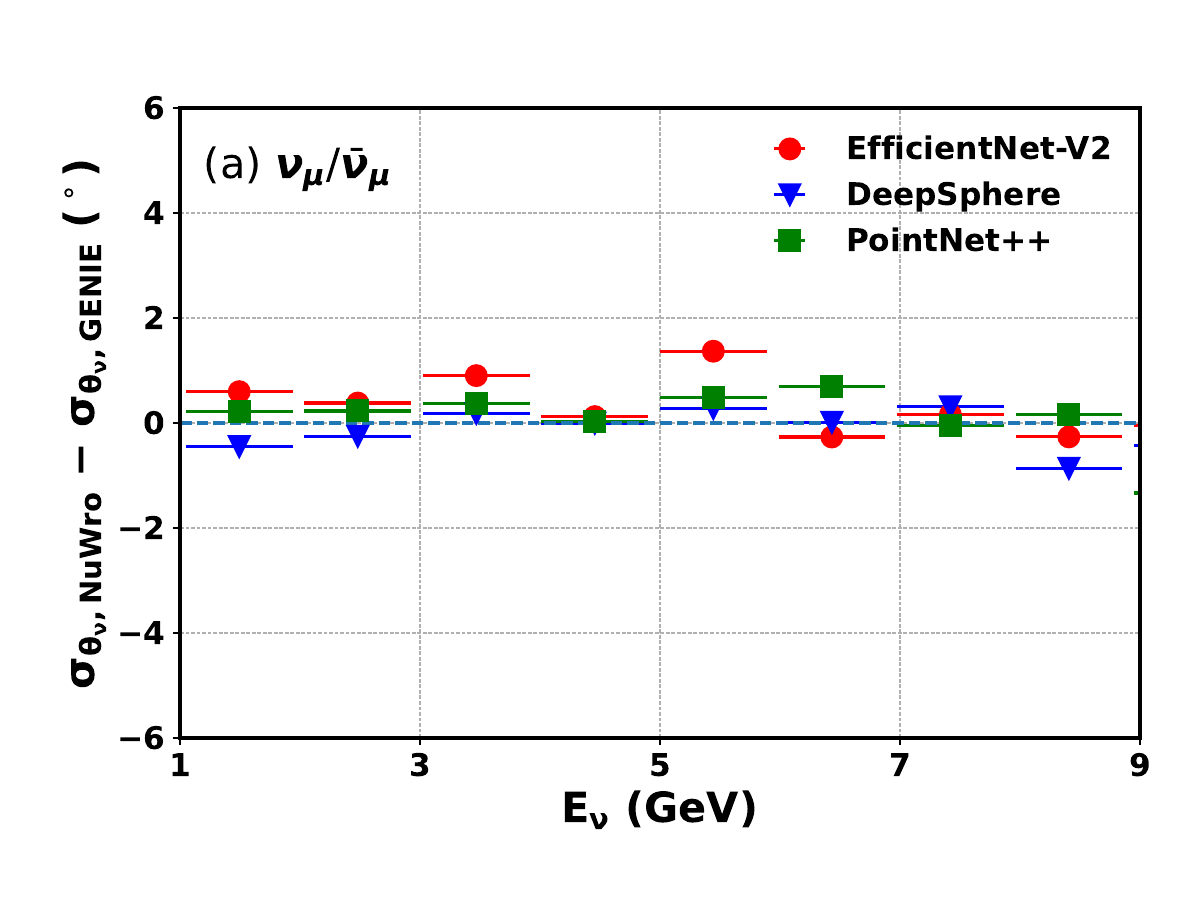}
\includegraphics[width=0.45\textwidth]{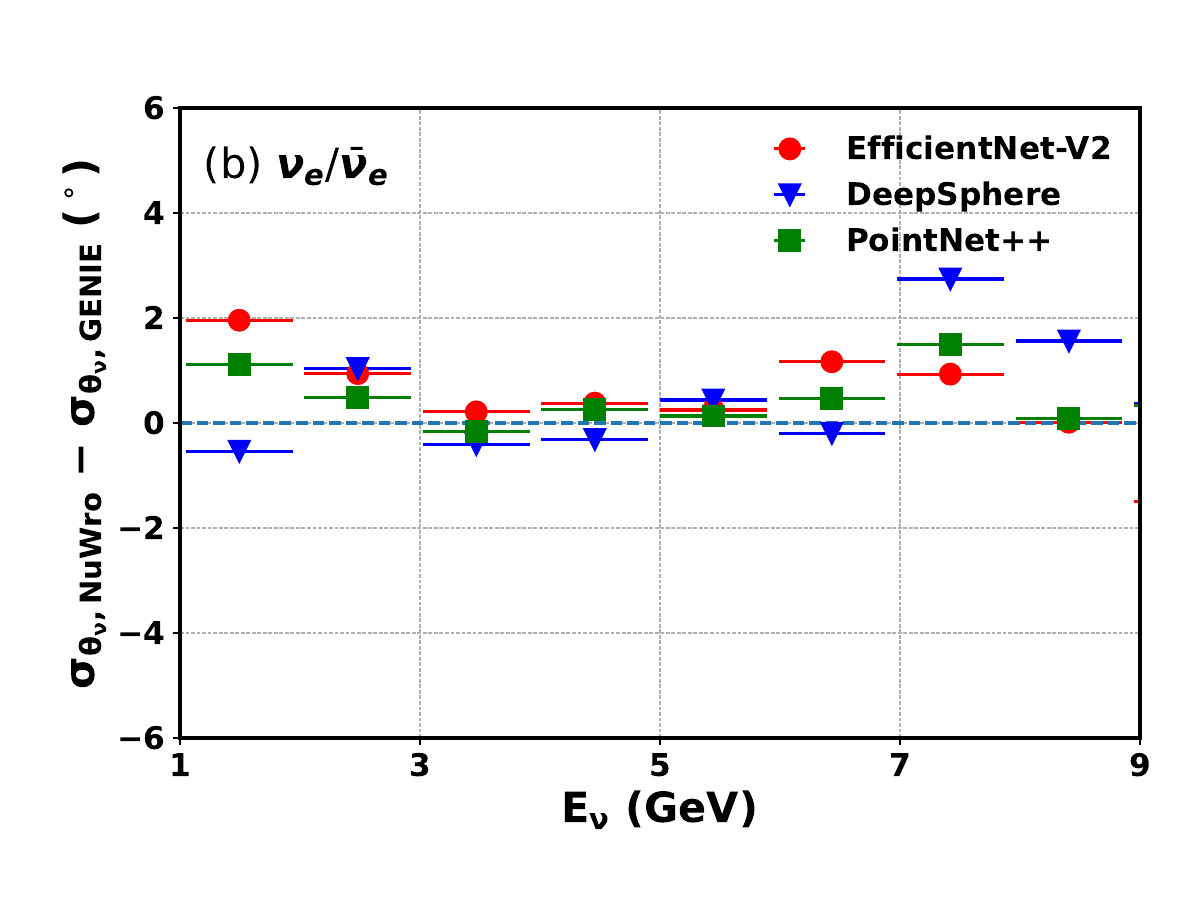}
\caption{\label{fig:genie_vs_nuwro } 
The difference in the $\theta_\nu$ resolutions obtained from (a) $\nu_\mu$/$\bar{\nu}_\mu$‐CC and (b) $\nu_e$/$\bar{\nu}_e$‐CC samples simulated by NuWro and GENIE using different ML models as functions of incoming neutrino energy. The ML models are trained on the same GENIE sample and tested on either an independent GENIE sample or a NuWro sample.  }
\end{figure*}

In this work, the incident atmospheric neutrino’s direction is reconstructed, which is different from most other atmospheric neutrino measurements where the final-state charged lepton's direction is used. 
This is a feasible strategy in an LS detector such as JUNO for two reasons: most importantly, both the charged lepton and hadrons can fully deposit their energy and produce scintillation light received by PMTs, preserving most of the incident neutrino's information;  
Furthermore, powerful ML models are able to resolve the complex relationship between the neutrino directionality and the PMT waveform features.
Fig.~\ref{fig:included_angles} compares the angle between the true and reconstructed neutrino directions, $\alpha$, using the PointNet++ result as an example, with the one between the incident neutrino and final-state charged lepton using the same $\nu_\mu$‐CC and $\nu_e$‐CC samples as described in section \ref{sec:sim}.  
The former is considerably smaller than the latter, by about $11^{\circ}$ if evaluated by the 68\% quantile, for both flavors.
This implies that the reconstructed neutrino direction is better than a perfectly measured final-state charged lepton direction in reflecting the true neutrino directionality, which can be an advantage for large LS detectors in atmospheric neutrino oscillation measurements. 
For water Cherenkov detectors, in comparison, hadrons produced in atmospheric neutrino interactions are often below the Cherenkov threshold and invisible, making it more difficult to reconstruct the incident neutrino direction.

While being able to utilize the information from hadrons brings an advantage, it can also potentially be subject to uncertainties from the neutrino interaction models used in the simulation.
To understand the dependence of this reconstruction method on neutrino interaction models, an independent sample simulated with an alternative generator, NuWro, is used. 
The ML models trained with the GENIE sample are used to reconstruct the events of the NuWro sample. 
The differences between the resolutions obtained from GENIE and NuWro samples as functions of neutrino energy are shown in Fig.~\ref{fig:genie_vs_nuwro }.
The results are consistent in most energy bins for all three models with the maximum difference around $2^\circ$, indicating that the method is robust and has marginal dependence on neutrino interaction models.

In addition to neutrino directionality, the PMT waveform can also be influenced by other event details, such as the track/shower starting/stopping points and energy deposition (d$E$/d$x$). 
Therefore, in principle, the method developed in this work by utilizing PMT waveform analysis and ML models can also be applied to the reconstruction of additional event information such as energy, interaction vertex, track trajectories, and particle types.
These tasks can be accomplished simply by adjusting the combination of input features and the output of the ML models.

\section{Summary and outlook \label{sec:so}}
This is the world’s first attempt to reconstruct atmospheric neutrinos' directionality in a large homogeneous LS detector. 
Despite their wide applications in various neutrino physics topics, such detectors have never been used for atmospheric neutrino oscillation measurements before. 
In this study, we demonstrate for the first time that an LS detector can offer good angular resolution for atmospheric neutrino oscillation measurements with waveform analysis and ML techniques. 
Different ML models and neutrino event generators are tested. The performance differences obtained are small and can be treated as systematic uncertainties. 
This method also has the advantage of reconstructing the neutrino direction directly rather than the final-state charged lepton direction, which can potentially further improve the neutrino oscillation sensitivity. 
Combined with good energy resolution, this work makes a large LS detector such as JUNO an excellent candidate for future atmospheric neutrino oscillation measurements.

\section*{Acknowledgements}
This work was partially supported 
by the National Natural Science Foundation of China (Grant No.12275160, No.12105158, No.12025502),  
by the Strategic Priority Research Program of the Chinese Academy of Sciences (XDA10010100), 
by CAS Project for Young Scientists in Basic Research (YSBR-099), 
by the Natural Science Foundation of Shandong Province (Grant No. ZR2022MA062) 
and by the China Postdoctoral Science Foundation (Project No. 2202M713153).  
We would also like to thank the 
Computing Center of the Institute of High Energy Physics,  Chinese Academy of Science, 
and the Key Laboratory of Particle Physics and
Particle Irradiation of Ministry of Education, Shandong University for providing the GPU resources.

\appendix

\bibliography{ref}

\end{document}